\documentclass[letterpaper,12pt]{article}

\usepackage{color}
\usepackage[normalem]{ulem}
\usepackage{amsmath,amssymb,slashed}
\usepackage[range-units=single,range-phrase=--]{siunitx}
\usepackage{bbm}
\usepackage{epsfig}
\usepackage{grffile,graphicx}
\usepackage{url}
\usepackage{enumerate}
\usepackage{multirow}
\usepackage{placeins}
\usepackage[dvipsnames]{xcolor}
\usepackage{epstopdf}
\usepackage[utf8]{inputenc}
\usepackage[version=4]{mhchem}
\usepackage{tablefootnote}
\usepackage[toc, acronym, nonumberlist]{glossaries}

\usepackage{tabularx} 
\usepackage{amsmath}  
\usepackage{graphicx} 
\usepackage[margin=1in,letterpaper]{geometry} 
\usepackage{cite} 
\usepackage[final]{hyperref} 
\hypersetup{
	colorlinks=true,       
	linkcolor=blue,        
	citecolor=blue,        
	filecolor=magenta,     
	urlcolor=blue         
}
\usepackage{amsmath,amssymb}
\usepackage{caption}
\usepackage{subcaption}
\usepackage{mathtools}
\usepackage{slashed}
\usepackage{bbm}
\usepackage{color}
\usepackage{xcolor}
\usepackage{graphics}
\usepackage{graphicx}
\usepackage{authblk}
\usepackage{tikz}
\newcommand{\be}{\begin{equation}}
\newcommand{\ee}{\end{equation}}
\newcommand{\beq}{\begin{equation}}
\newcommand{\eeq}{\end{equation}}
\newcommand{\bea}{\begin{eqnarray}}
\newcommand{\eea}{\end{eqnarray}}

\newcommand{\Tend}{T_\text{end}}
\newcommand{\Teq}{T_\text{eq}}
\newcommand{\Tosc}{T_\text{osc}}
\newcommand{\tosc}{t_\text{osc}}
\newcommand{\Tc}{T_\text{c}}
\newcommand{\Rend}{R_\text{end}}
\newcommand{\Req}{R_\text{eq}}
\newcommand{\Rosc}{R_\text{osc}}
\newcommand{\Rc}{R_\text{c}}
\newcommand{\Heq}{H_\text{eq}}
\newcommand{\rhoeq}{\rho_\text{eq}}

\newcommand{\gs}{g_\star}
\newcommand{\gss}{g_{\star S}}

\newcommand{\rp}{\rho_\phi}
\newcommand{\rR}{\rho_R}

\newcommand{\TBBN}{T_\text{BBN}}

\newcommand{\fa}{f_a}
\newcommand{\maT}{\tilde m_a}
\newcommand{\ma}{m_a}

\newcommand{\Omegastdo}{\Omega_{\rm std,-3/2}}
\newcommand{\OmegastdT}{\Omega_{\rm std, -7/6}}

\newcommand{\Toscthree}{\Tosc^{R_3}}

\newcommand{\TQCD}{T_{\rm QCD}}
\newcommand{\ReportNumbers}[1]{%
\begin{tikzpicture}[overlay, remember picture]
\path (current page.north east) ++(-1,-1) node[below left] {#1};
\end{tikzpicture}
}

\newcommand{\lrb}[1]{\left( #1 \right)}

\begin{document}
\begin{center}
\vspace*{0.5cm}
{\Large{\textbf{
New opportunities for axion dark matter searches\\in nonstandard cosmological models
}}}
\date{}

\vspace*{1.2cm}

{\large
Paola Arias$^{a}$,\,%
Nicol\'as Bernal$^{b}$,\,%
Dimitrios Karamitros$^{c}$,\\%
Carlos Maldonado$^{a}$,\,%
Leszek Roszkowski$^{c,d}$\,%
and Moira Venegas$^{a}$
}\\[3mm]
{\it{
$^{a}$ Departamento de Física, Universidad de Santiago de Chile, Casilla 307, Santiago, Chile
\\
$^{b}$ Centro de Investigaciones, Universidad Antonio Nariño\\Carrera 3 Este \# 47A-15, Bogotá, Colombia
\\
$^{c}$ National Centre for Nuclear Research, ul. Pasteura 7, 02-093 Warsaw, Poland
\\
$^{d}$ 
AstroCeNT, Nicolaus Copernicus Astronomical Center Polish Academy of Sciences,\\
ul. Rektorska 4, 00-614 Warsaw, Poland
}}

\end{center}

\ReportNumbers{\footnotesize PI/UAN-2021-693FT}

\begin{abstract}
We study axion dark matter production from a misalignment mechanism in scenarios featuring a general nonstandard cosmology.
Before the onset of Big Bang nucleosynthesis, the energy density of the universe is dominated by a particle field $\phi$ described by a general equation of state $\omega$.
The ensuing enhancement of the Hubble expansion rate decreases the temperature at which axions start to oscillate, opening this way the possibility for axions heavier than in the standard window.
This is the case for kination, or in general for scenarios with $\omega > 1/3$.
However, if $\omega < 1/3$, as in the case of an early matter domination, the decay of $\phi$ injects additional entropy relative to the case of the standard model, diluting this way the preexisting axion abundance, and rendering lighter axions viable.
For a misalignment angle $0.5 < \theta_i < \pi/\sqrt{3}$, the usual axion window becomes expanded to $4 \times 10^{-9}$~eV $\lesssim m_a \lesssim 2 \times 10^{-5}$~eV for the case of an early matter domination, or to $2 \times 10^{-6}$~eV $\lesssim m_a \lesssim 10^{-2}$~eV for the case of kination.
Interestingly, the coupling axion-photon in such a wider range can be probed with next generation experiments such as ABRACADABRA, KLASH, ADMX, MADMAX, and ORGAN.
Axion dark matter searches may therefore provide a unique tool to probe the history of the universe before Big Bang nucleosynthesis.
\end{abstract}
\thispagestyle{empty} 

\newpage
\tableofcontents

\section{Introduction}  
The axion is a by-product of an elegant solution of the strong CP problem of the QCD sector~\cite{Peccei:1977hh, Weinberg:1977ma, Wilczek:1977pj} of the standard model of particle physics (SM). Interestingly,  the same mechanism that solves the strong CP problem, leads to an efficient mechanism of nonthermal production of a cold  population in the early universe, making the axion an excellent cold dark matter (CDM) candidate~\cite{Preskill:1982cy, Dine:1982ah, Abbott:1982af}. Axions appear as pseudo-Nambu-Goldstone bosons when the so-called Peccei-Quinn symmetry (PQS) is spontaneously broken at some energy $f_a \gg v_{\rm EW} \simeq 247$~GeV. As the universe cools down, the axion potential energy changes during the QCD phase transition epoch, due to instanton effects that break explicitly the PQS, acquiring a small mass. During this process, the value of the axion field gets realigned (the process is known as the 
``misalignment mechanism''), changing from an arbitrary initial value $a_i$, to the true vacuum value of the field, $\langle a\rangle =0$. This process is of extreme relevance, since on the one hand, it solves the so-called strong CP problem of the QCD sector and fills the universe with cold axion particles produced during the oscillation of the field around the minimum.

A distinctive feature of the axion is that its mass has a definite relationship with the PQ scale  by $f_a\, \maT(T) = \sqrt{\chi(T)}$, where $\chi$ is the topological susceptibility in QCD, which has been evaluated in the chiral limit~\cite{Crewther:1977ce, DiVecchia:1980yfw}, next-to-next-to leading order in chiral perturbation theory~\cite{Gorghetto:2018ocs} and directly via QCD lattice simulations~\cite{Borsanyi:2016ksw}, all coinciding with the central limit
\be \label{eq:ma}
    \ma \simeq 5.69\,\mbox{meV} \left(\frac{10^{9}\,\mbox{GeV}}{f_a} \right)\, ,
\ee
where $\ma$ corresponds to the  axion mass at zero temperature.
However, for temperatures higher than the one of the QCD phase transition, the axion mass is suppressed.

Only two parameters are needed to fix the relic density of axions: the scale $f_a$ (or equivalently its mass) and the initial value of the field at the moment when it starts to oscillate, $a(\Tosc)\simeq a_{i}$.
In a pre-inflationary scenario, the PQS breaking occurs before or during inflation and it is not restored afterwards. In this scenario, the axion field takes different initial values in different disconnected patches that are eventually stretched during inflation, homogenising its initial value in our universe to $a_i=f_a\, \theta_i$, where the initial angle is randomly selected in the range $\theta_i \in \left[-\pi, \pi\right]$. For $\theta_i\sim \mathcal O(1)$, $\ma$ should be in the range of $\mu$eV in order for the axion density not to exceed the observed CDM density $\Omega_{\rm CDM}h^2 \simeq 0.12$~\cite{Bae:2008ue, Wantz:2009it, Borsanyi:2016ksw, Ballesteros:2016xej, Planck:2018vyg}. Smaller masses (equivalently, higher $\fa$ scales) can be reached if small-tuned values of $\theta_i$ are considered. This is the so-called anthropic axion window~\cite{Hertzberg:2008wr, Wantz:2009it}. Nonetheless, in this pre-inflationary scenario, high values of $\fa$ are constrained by isocurvature perturbations~\cite{Beltran:2006sq, Hertzberg:2008wr, Planck:2018vyg}, requiring $\fa \lesssim 10^{16}$~GeV.  On the other hand, in a post-inflationary scenario, the PQS breaks after inflation, so $\theta_i$ takes different values in different patches of the present universe, so its value gets averaged to $\bar \theta_i= \pi/\sqrt{3}$, which then sets the DM abundance to be correctly satisfied for $\ma \simeq 30~\mu$eV. However, in this scenario topological defects such as cosmic strings and domain walls emerge, contributing to the CDM abundance in a still disputed quantity due to the uncertainties in their energy loss process~\cite{Hagmann:2000ja, Wantz:2009mi, Hiramatsu:2010yu, Kawasaki:2014sqa, Gorghetto:2018ocs}. Thus, in the post-inflationary PQS breaking the accepted axion mass is in the range $30~\mu$eV $\lesssim \ma \lesssim 5$~meV.

The QCD axion has been realised in many scenarios, where the most popular ones correspond to the DFSZ model~\cite{dine1981simple, Zhitnitsky:1980tq} and the KSVZ model~\cite{kim1979weak, shifman1980can, DiLuzio:2016sbl}. The axion can feature couplings to nucleons, electrons and more importantly to two photons. Stellar evolution arguments  constrain the previously mentioned interactions, where  the most stringent one comes from the coupling of axions to nucleons, leading to an axion flux from a supernova explosion that  otherwise would quench the observed neutrino pulse from SN1987a, unless the axion decay constant is constrained to be~\cite{Ellis:1987pk, Raffelt:1987yt, Turner:1987by}
\be
    \fa\gtrsim 4 \times 10^8~\mbox{GeV}, 
\ee
corresponding to a mass of $\ma\lesssim 1.6\times 10^{-2}$~eV. 

The uncertainty in the astrophysical models and assumptions on axion coupling to fermions makes laboratory based searches crucial to test axions in the  $\mu$eV mass ballpark. So far, only the so-called haloscope experiments~\cite{Sikivie:1983ip}, ADMX~\cite{ADMX:2009iij, ADMX:2019uok} and HAYSTAC~\cite{HAYSTAC:2018rwy}, using a highly tuned microwave cavity that converts DM axions into photons in the presence of a static magnetic field, have achieved enough sensitivity to touch the axion band in the sub-eV regime.  Several operating and proposed haloscopes aim to explore  masses in between the $\mu$eV to the meV range, such as  CULTASK~\cite{CAPP:2020utb, Lee:2020cfj} KLASH~\cite{Alesini:2017ifp}, ORGAN~\cite{McAllister:2017lkb, McAllister:2020twv}, RADES~\cite{Melcon:2018dba, CAST:2021add} and QUAX~\cite{Alesini:2020vny}. A related concept makes use of dielectric plates, such as the MADMAX~\cite{MADMAX:2019pub} experiment. The LC circuit based detection~\cite{Sikivie:2013laa}  is expected to reach axion masses in the $10^{-7}$ to $10^{-9}$~eV range. The ABRACADABRA experiment~\cite{Kahn:2016aff,Ouellet:2018beu} and ADMX SLIC~\cite{Crisosto:2019fcj} have already released promising results. Topological insulators have the potential to explore masses in the few~$\times$~meV range \cite{Schutte-Engel:2021bqm}. In a higher mass range, the helioscope experiment CAST has explored the $\ma\sim $~eV ballpark and it is expected to improve with the next generation experiment IAXO~\cite{IAXO:2019mpb}. For details on axion searches, prospects, and their schedules, we refer, e.g., to Refs.~\cite{Sikivie:2020zpn, Irastorza:2018dyq, Billard:2021uyg}. 

So far, no signal of axion DM has been found, but given the many forthcoming experiments that will access to probe the axion CDM prediction, it is expected that in the following years either a discovery is made, or the axion is not found in that parameter space. In the second pessimistic scenario, there are several ways to go around the preferred DM parameter space, such that the axion is either even more weakly  or strongly coupled than thought. The most straightforward approach is to consider that in the case of the pre-inflationary scenario, there is a very small initial misalignment  angle $\theta_i$, to compensate for a higher $f_a$ scale shifting it  to be as high as $\sim 10^{16}$~GeV, as allowed by bounds on isocurvature perturbations. Another way to open up the axion DM window to smaller masses or higher PQ scales invokes the coupling of the axion to some other field, for instance, to a hidden photon~\cite{Agrawal:2017eqm, Agrawal:2018vin}, or if the axion potential was much larger in the very early universe~\cite{Choi_1997, Banks_1997, Banks_2003, Heurtier:2021rko}. A different possibility to expand the range masses has been to consider a nonstandard cosmological (NSC) history in the early universe, such that  entropy is injected into the thermal bath, diluting the axion energy density.

In this work we pursue the last alternative, that is, consider a NSC history of the early universe prior to Big-Bang Nucleosynthesis (BBN) by means of a new field that eventually dominates the expansion of the universe. 
There are several works that have previously studied a NSC scenario in the context of axion physics. Besides, pioneer papers that considered an early matter dominance (EMD) period~\cite{Steinhardt:1983ia, Lazarides:1990xp, Kawasaki:1995vt, Giudice:2000ex}, in Ref.~\cite{Visinelli:2009kt, Venegas:2021wwm} the authors studied the cases of low temperature reheating (LTRH) and kination, including the anharmonicities in the axion potential. In Refs.~\cite{Grin:2007yg, Carenza:2021ebx} thermally produced axions in LTRH and kination cosmologies were analysed, whereas in Ref.~\cite{Blinov:2019jqc} it was considered the misalignment production of axion-like particles in early matter domination and kination cosmologies. In Ref.~\cite{Ramberg:2019dgi} a full scan of cosmological histories was performed, together with including the contribution from the decay of an axionic string network. In Refs.~\cite{Nelson:2018via, Visinelli:2018wza} the impact of NSC on the formation of axion miniclusters was analysed.
Finally, the axion DM scenario in a NSC induced by a primordial black hole domination era was recently studied in Ref.~\cite{Bernal:2021yyb}.

Our approach is to consider a very detailed analysis of the DM production during a NSC scenario. In particular, we study analytically the axion relic density and we find the mass that can account for the whole DM abundance observed today as a function of the NSC parameters. We also solve numerically to check our results, getting a good agreement. We find the features and characteristics of the NSCs where the misalignment mechanism can lead to an extended range in the axion mass and we map their impact on the coupling to two-photon plane. We point the experimental efforts and prospects that can test those cosmological scenarios.

The manuscript is organised as follows: in the next section~\ref{sec:misalignment_std} we will review the standard misalignment mechanism for axions in the framework of a standard cosmological scenario. We will derive the analytical expressions of the relic density and compare them with the numerical results. Then in section~\ref{sec:nsc} we introduce the characteristics and parameters that define the models of NSC considered. In section~\ref{sec:nsc_oscillation}  we derive the axion relic density in terms of the NSC parameters and then we contrast with the numerical results. We divide our analysis,  starting with cosmologies with an equation of state $\omega<1/3$, where we choose as benchmark values  $\omega=0$ and $\omega=-1/3$. Then, we study the case $\omega>1/3$, using as benchmarks $\omega=1$ and $\omega=5/3$. In section \ref{sec:axion_coupling} we map the results into the exclusion plot for the coupling of axions to two photons. We show the opportunities to extend the DM parameter space for different cosmologies, and we mention on the new generation of experiments that will be able to test some of that parameter space. Finally, in section~\ref{sec:conclusions} we summarise and conclude. 

\section{Axion DM in the standard cosmological scenario: misalignment essentials} \label{sec:misalignment_std}

Let us start by reviewing the essentials of the misalignment production and write down the parameter space where the axion can explain the whole DM observed today. 
The temporal evolution of the SM temperature $T$ can be obtained from the evolution of the SM entropy density $s(T) = \frac{2\pi^2}{45}\,\gss\,T^3$ given by
\begin{equation} \label{eq:entropySC}
    \frac{ds}{dt} + 3\, H\, s = 0\,,
\end{equation}
where $\gss(T)$ accounts for the number of relativistic degrees of freedom present in the SM entropy~\cite{Drees:2015exa}, and $H$ corresponds to the Hubble expansion rate which, in a radiation dominated universe, is given by
\be \label{eq:hubble_std}
    H(T) = \sqrt{\frac{\rho_R(T)}{3\, M_P^2}} = \frac{\pi}{3} \sqrt{\frac{\gs(T)}{10}}\, \frac{T^2}{M_P}\,,
\ee
where $\rho_R(T)$ is the SM radiation energy density, $\gs(T)$ accounts for the relativistic degrees of freedom contributing to $\rho_R$~\cite{Drees:2015exa}, and $M_P \simeq 2.4\times 10^{18}$~GeV is the reduced Planck mass.
Equation~\eqref{eq:entropySC} allows to extract the temperature evolution in terms of the scale factor $R$ as
\begin{equation} \label{eq:dTdR}
    \frac{dT}{dR} = - \left(1+\frac{T}{3\,\gss}\frac{d\gss}{dT}\right)^{-1} \left(\frac{T}{R}\right).
\end{equation}

Furthermore, the scale $f_a$ at which the PQS is spontaneously broken determines the axion mass through the topological susceptibility of QCD, $\chi(T)$, as 
\be 
    \maT^2(T) = \frac{\chi(T)}{f_a^2}\,,
\ee
where $\chi(T)$ has been estimated from lattice QCD simulations and found a zero-temperature value of $\chi_0  \equiv \chi(0) \simeq 0.0245$~fm$^{-4}$, in the symmetric isospin case~\cite{Borsanyi:2016ksw}. For the numerical calculations the results of Ref.~\cite{Borsanyi:2016ksw} will be used, however, for analytical estimations we will take instead an approximate expression, which has to be cut off by hand once the mass reaches the zero-temperature value~\cite{Hertzberg:2008wr}
\be \label{eq:thermal_mass}
\tilde m_a(T) \simeq \ma \times
    \begin{dcases}
        1 &\text{ for } T \leq \TQCD\,,\\
        \left(\frac{T}{\TQCD}\right)^{-4} &\text{ for } T \geq \TQCD\,,
    \end{dcases}
\ee
with $\ma$ given in Eq.~\eqref{eq:ma}, and $\TQCD \simeq 150$~MeV the approximate temperature at the QCD phase transition.
We have checked that this approximation closely follows the lattice results from Ref.~\cite{Borsanyi:2016ksw}.

To track the evolution of the axion field during the early universe, let us write down the axion Lagrangian density as
\be \label{axion_lag}
    \mathcal L = \frac12\, \partial_\mu a\, \partial^\mu a-\maT^2(t)\, f_a^2 \left(1-\cos\frac{a}{f_a}\right).
\ee
For an homogeneous axion field, we only track down the evolution of the zero mode, which is found to be
\be \label{axion_eom}
    \ddot \theta + 3\, H(t)\, \dot\theta + \maT^2(t)\, \sin \theta = 0\,,
\ee
where $\theta(t) \equiv a(t)/f_a $.
At high energies ({\it i.e.}, $T \gg \TQCD$), the last term can be ignored and the field is stuck on a constant value $\theta_{i}$, which is in principle randomly selected.
At temperature $T = \Tosc$, defined as the temperature where the equality
\begin{equation}
    3\, H(\Tosc) = \maT(\Tosc)    
\end{equation}
holds, the axion field starts to oscillate around its true minimum, $\theta = 0$.%
\footnote{We note that there is a degree of arbitrariness in the definition of $\Tosc$, as one can also have $A\, H(\Tosc) = \maT(\Tosc)$, with $A$ typically being between 1 and 3, see e.g. Refs.~\cite{Marsh:2015xka, Blinov:2019rhb}. Here however we fix $A=3$ as it gives a good fit to the full numerical result.}
Within the WKB approximation, {\it i.e.} $\theta \ll 1$ and slowly varying $H$
and $\maT$ (adiabatic evolution of the axion),%
\footnote{For larger values of the angle, one has to include the ``anharmonic" contribution to the potential. In our numerical results, the anharmonic factor derived in Appendix~\ref{app:anF} are included.} 
\be \label{eq:axion_field_WKB}
a(t) = a_i  \left[\frac{\maT(\tosc)}{\maT(t)} \left(\frac{\Rosc}{R(t)}\right)^3\right]^{1/2} \cos\lrb{\int \maT(t)\, dt}.
\ee
We should note that this we also have assumed that the value of the field at $t=\tosc$ is close to its initial value, {\it i.e.}  $a(t=t_{\rm osc}) \simeq a_i$. 
It is convenient to rewrite Eq.~\eqref{axion_eom} in terms of the scale factor as
\be \label{eq:angle_R}
    \theta'' + \left(\frac4R + \frac{H'(R)}{H(R)}\right)\theta' + \left(\frac{\maT(R)}{H(R)\, R}\right)^2 \sin\theta = 0\,,
\ee
where primes denote derivatives with respect to $R$.

\begin{figure}[t]
    \centering
    \includegraphics[width=0.5\textwidth]{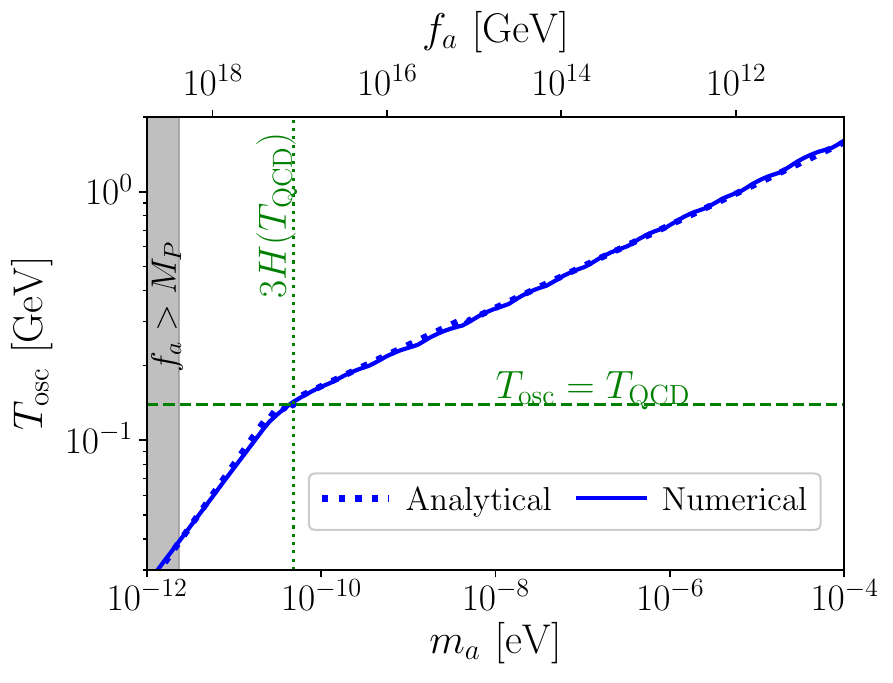}
    \includegraphics[width=0.49\textwidth]{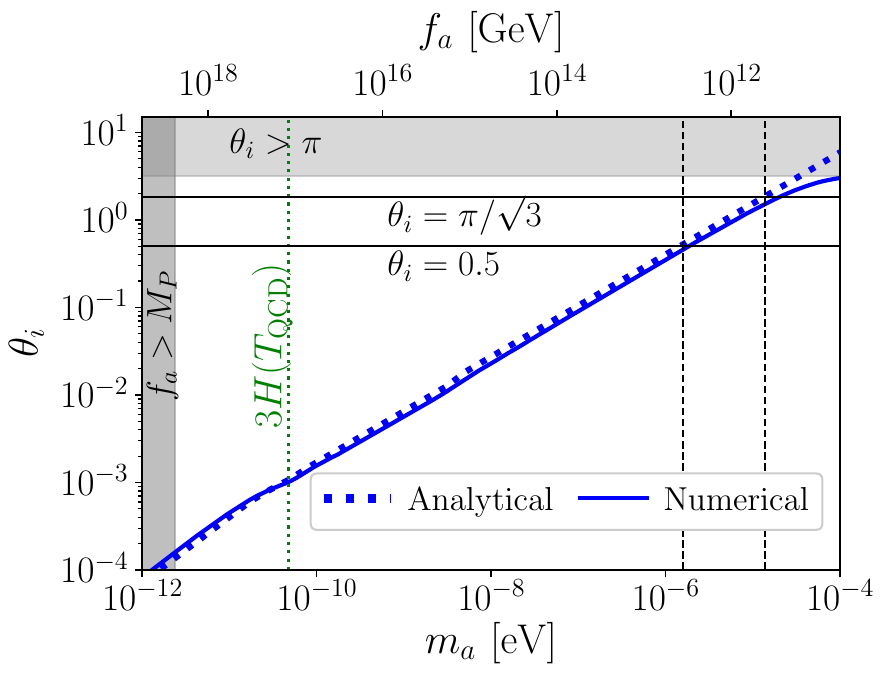}
    \caption{Left panel: Oscillation temperature as function of the axion mass. The bend around $\ma \simeq 4.8 \times 10^{-11}$~eV corresponds to the QCD transition.
Right panel: Initial misalignment angle as a function of the axion mass. In both panels, the blue solid line is for the numerical solution and dotted blue line is for the analytical solution. The mismatch for $\theta_i>\pi/\sqrt{3}$ is due to the fact that we do not consider anharmonic effect in the analytical solution. In both panels the dark grey areas correspond to the scale $f_a$ bigger than the Planck scale.}
    \label{fig:axion_std}
\end{figure}

The axion energy density $\rho_a$ is given by
\be\label{eq:rho_a_t}
    \rho_a(t) = \frac{\dot a^2}{2} + \maT^2(t)\, \fa^2 \left(1-\cos\frac{a}{f_a}\right)
    \simeq \frac{\dot a^2}2+\frac{\maT^2(t)\, a^2}{2}\,,
\ee
where in the last step $a/f_a \ll 1$ was assumed.
It is interesting to note that even if $\rho_a$ is not conserved since its mass (and the scale factor) vary with time, its comoving number $N_a \equiv n_a\, R^3$ is conserved. The energy density for non-relativistic axions is $\rho_a(t)=\maT(t)\, n_a(t)$, and at present ({\it i.e.}, at $t = t_0$) becomes
\be \label{axion_density_sc}
    \rho_a(t_{0}) = \rho_a(\tosc)\, \frac{\ma}{\maT(\tosc)} \left(\frac{R(\tosc)}{R_0}\right)^3
    \simeq \frac{\theta_i^2}{2}\, f_a^2\, \ma\, \maT(\Tosc)\, \frac{s(T_0)}{s(\Tosc)}\,,
\ee
with $R_0 \equiv R(t_0)$ the scale factor at present, and using entropy conservation and $\rho_a(\Tosc) \simeq \frac12\, \maT^2(\Tosc)\, f_a^2\, \theta_i^2$.
The oscillation temperature $\Tosc$ is therefore
\bea \label{eq:Tosc_std}
    \Tosc \simeq
    \begin{dcases}
        \left(\frac{1}{\pi} \sqrt{\frac{10}{\gs(\Tosc)}}\, \ma\, M_P\right)^{1/2} &\text{for }\,\Tosc\leq \TQCD\,,\\
        \left(\frac{1}{\pi} \sqrt{\frac{10}{\gs(\Tosc)}}\, \ma\, M_P\, \TQCD^4\right)^{1/6} &\text{for }\,\Tosc\geq \TQCD\,.
    \end{dcases}
\eea
The dependence of $\Tosc$ with $\ma$ is depicted in the left panel of Fig.~\ref{fig:axion_std}, where the change in the slope at $\Tosc = \TQCD$ is shown with a horizontal line.  
The corresponding axion mass is
\be \label{eq:m_i}
    3\, H(\TQCD) = \pi\, \sqrt{\frac{\gs(\TQCD)}{10}}\, \frac{\TQCD^2}{M_P} \simeq 4.8 \times 10^{-11}~\text{eV}.
\ee
Thus, for masses above $3\, H(\TQCD)$, the temperature effects of QCD are important, and below they can be safely ignored. 
Now we can insert $\Tosc$ back into Eq.~\eqref{axion_density_sc} in order to have the axion relic density today
\be
    \Omega_a \equiv \frac{\rho_a(T_0)}{\rho_c} \simeq
    \begin{dcases}
        0.006 \left(\frac{\theta_{i}}{1}\right)^{2}\left(\frac{\ma}{5.6~\mu \mbox{eV}}\right)^{-3/2} &\text{ for } \ma \lesssim 3\, H(\TQCD)\,,\\
        0.17 \left(\frac{\theta_{i}}{1}\right)^{2}\left(\frac{\ma}{5.6~\mu \mbox{eV}}\right)^{-7/6} &\text{ for } \ma \gtrsim 3\, H(\TQCD)\,,
    \end{dcases}
    \label{eq:relic_stdd}
\ee
where we have used the fact that $\rho_c \simeq 1.1 \times 10^{-5}~h^2$~GeV/cm$^3$ is the critical energy density, $h \simeq 0.674$, and $s(T_0) \simeq 2.9 \times 10^3$~cm$^{-3}$~\cite{Planck:2018vyg}.
In order to match the observed DM relic abundance $\Omega_a \simeq 0.26$~\cite{Planck:2018vyg}, the initial misalignment angle $\theta_i$ could be tuned.
The right panel of Fig.~\ref{fig:axion_std} shows the required angle $\theta_i$ in order to generate the whole observed DM relic abundance, for different axion masses. It is worth pointing-out that the analytical approximation deviates significantly from the numerical result for $\theta_i \gtrsim 1$, since the former uses the exact potential as well as the corresponding anharmonic factor derived in Appendix~\ref{app:anF}.

The initial angle is deeply connected to the cosmological history of the axion: the realignment of the axion potential happens in different causally disconnected patches in the universe with (in principle) different initial misalignment angles.  If the PQS is broken before or during inflation (and not restored afterwards) one of these patches gets inflated away, homogenising the initial angle over the observable universe and the angle is randomly selected  from $\left[-\pi, \pi\right]$. On the other hand, if the PQS breaking occurs after the inflationary epoch,  different initial angles in the different disconnected patches are arbitrarily selected between  $\left[-\pi, \pi\right]$, thus, the angle is averaged to $\langle \theta_i^2\rangle = \pi^2/3 \simeq (1.8)^2$~\cite{Hertzberg:2008wr}. Therefore, the parameter space where the axion can explain the DM paradigm can be different, depending on whether $f_a$ is higher or lower than  $T_I \equiv H_I/2\pi$, where $H_I$ corresponds to the inflationary scale.
For the pre-inflationary scenario, where $f_a\geq T_I$, there is a strong constraint on high $f_a$ scales from isocurvature perturbations~\cite{Hertzberg:2008wr}. If the axion is present during inflation, it develops quantum fluctuations which are nearly scale invariant and uncorrelated with the adiabatic perturbations. Using Planck constraints on the primordial isocurvature
fraction~\cite{Planck:2018vyg}, it is found that~\cite{Zyla:2020zbs}
\be
    H_I \lesssim 1.3\times 10^9~\mbox{GeV}  \left(\frac{\fa}{10^{16}\, \mbox{GeV}}\right)^{0.42},
\ee
assuming axions make the whole DM abundance.
On the other hand, an axion with a PQ scale $\fa \simeq 10^{16}$~GeV can produce the whole abundance for a misalignment angle of $\theta_i \simeq 3 \times 10^{-3}$ (cf. Fig.~\ref{fig:axion_std}), which is considered tuned or anthropic. That is, a pre-inflationary axion hints to a low-scale inflation in the standard cosmological scenario.

In the  post-inflationary scenario,  $\fa<T_I$, the bound on the PQ scale comes from axion DM not overclosing the universe, which requires $\fa \lesssim 10^{11}~$GeV~\cite{Wantz:2009mi}. On the other hand, in this scenario topological defects such as cosmic strings and domain walls emerge, radiating cold axions. Their contribution it is still in dispute \cite{Hagmann:2000ja, Wantz:2009mi, Hiramatsu:2010yu, Kawasaki:2014sqa, Gorghetto:2018ocs, Harari:1987us}, nonetheless it is usually assumed they  enlarge the DM parameter space down to   $\fa \gtrsim10^9$~GeV, translated into a mass range of approximately $56~\mu\mbox{eV}\lesssim \ma\lesssim 5$~meV.

For the rest of our discussion -- since we are only focused on studying the impact of NSC on the misalignment mechanism -- we will consider as the standard cosmology axion mass range, the one obtained from the misalignment mechanism, Eq.~\eqref{eq:relic_stdd}, by considering $\theta_i$ in a natural range of $\left[0.5, \pi/\sqrt{3}\right]$, that leads to masses $1.6\times 10^{-6}$~eV $\lesssim \ma \lesssim 1.4\times 10^{-5}$~eV.

\section{Nonstandard cosmologies} \label{sec:nsc}
It is typically assumed that the energy density of the universe was dominated by radiation between the end of the reheating era after inflation and the onset of matter domination at redshift $z \simeq 3400$, {\it i.e.}, the standard cosmological scenario.
However, there is no evidence that the universe was radiation dominated before the BBN, at $\TBBN \simeq 4$~MeV~\cite{Kawasaki:1999na, Kawasaki:2000en, deSalas:2015glj, Hasegawa:2019jsa}. 
In that context, NSCs have been widely studied, especially in the context of a fluid with an equation-of-state parameter $\omega$, with $-1 \leq \omega \leq 1$, that eventually dominates the energy density of the universe.%
\footnote{Has been argued that cosmologies with $\omega>1$ could feature superluminal propagation, although in Ref.~\cite{DEramo:2017gpl} they show that is not the case. Moreover, as discussed in Refs.~\cite{Allahverdi:2018iod, Allahverdi:2019jsc, Arias:2020qty}, scenarios with $\omega\to-1$ are disfavoured. A more concrete bound on $\omega$ (and the other parameters related to $\phi$) would depend on the underlying inflationary model, which is beyond our scope.}
Extensively studied NSCs are the EMD, firstly explored in the context of supersymmetry and string theory (see e.g. Refs.~\cite{Vilenkin:1982wt, Coughlan:1983ci, Moroi:1999zb}), with $\omega = 0$, and the kinetic energy domination, known as kination dominance, with $\omega = 1$~\cite{Barrow:1982ei, Ford:1986sy, Spokoiny:1993kt}.  For a detailed review on NSC, see Ref.~\cite{Allahverdi:2020bys}. In this work, we will consider the existence of a particle field $\phi$ with a general equation of state, $\omega = p_\phi / \rho_\phi$, that during a certain period  has an impact on the expansion of the universe. Prior to BBN, the field could decay into SM radiation, with a total decay rate $\Gamma_\phi$. The Boltzmann equations that govern the evolution of the energy density $\rho_\phi$ of $\phi$ and the SM entropy density $s$ are%
\footnote{It is interesting to note that, in the case of $\omega > 1/3$, $\rho_\phi$ gets diluted faster than radiation, and therefore $\phi$ could be stable, with $\Gamma_\phi = 0$.}
\begin{align}
    \frac{d\rp}{dt}+3(1+\omega)\,H\,\rp &=-\Gamma_\phi\,\rp\,,\label{eq:cosmo2} \\
    \frac{ds}{dt}+3\,H\,s &=+\frac{\Gamma_\phi}T\,\rp\,.\label{eq:cosmo3}
\end{align}
The latter equation can be recast to find the relationship between the plasma temperature $T$ and scale factor $R$ in the case of an entropy injection due to the decay of $\phi$, as
\begin{equation}\label{eq:cosmo3b}
    \frac{dT}{dR}=\left(1+\frac{T}{3\,\gss}\frac{d\gss}{dT}\right)^{-1}\left[-\frac{T}{R}+\frac{\Gamma_\phi\,\rp}{3\,H\,s\,R}\right].
\end{equation}
The Hubble expansion rate is given by
\be \label{eq:Hubble}
    H=\sqrt{\frac{\rho_\phi+\rho_R+\rho_a}{3M_P^2}}\,,
\ee
featuring the contribution from $\phi$.
As the axion contribution is always subdominant, the evolution of the $\phi$-radiation system is decoupled from the DM evolution and can be solved separately.
It is customary to define the end of the $\phi$ domination with the temperature $\Tend$.
For $\omega > 1/3$, as $\phi$ does not have to decay, $\Tend$ corresponds to the equality $\rho_R(\Tend) = \rho_\phi(\Tend)$.
However, for $\omega < 1/3$, it corresponds to the temperature at which the field has mostly decayed away, {\it i.e.} when $H(\Tend) = \Gamma_\phi$, and therefore~\cite{Chung:1998rq, Giudice:2000ex}
\begin{equation}\label{eq:Tend}
    \Tend^4\equiv\frac{90}{\pi^2\,\gs(\Tend)}\,M_P^2\,\Gamma_\phi^2\,.
\end{equation}

\begin{figure}[t]
\center
\includegraphics[scale=0.84]{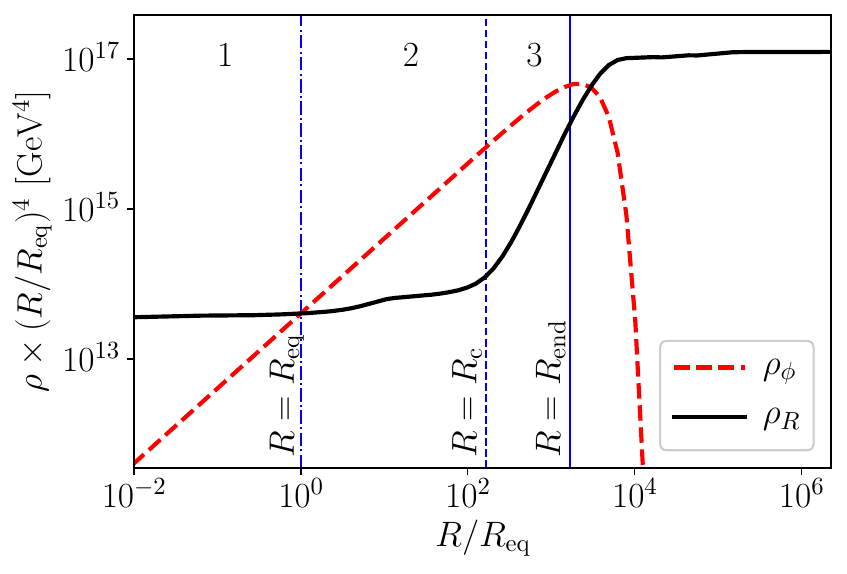}
\caption{Evolution of the energy densities for radiation (black) and  $\phi$ field (dashed red) as a function of the scale factor $R$, for $\beta = 3$, $\Teq =1$~GeV, and $\Tend = 4$~MeV. The vertical lines depict $R = R_\text{eq}$ (solid blue), $R = R_c$ (dashed blue), and $R = R_\text{end}$ (dash-dotted line), respectively.}
\label{fig:energies_nsc}
\end{figure}

Equations~\eqref{eq:cosmo2} and~\eqref{eq:cosmo3} can be analytically solved.
Here we present simplified approximations that will be used in our analytical analysis later.
However, the complete details are outlined in the Appendix~\ref{app:phi-r}.
Before the decay of $\phi$, the energy densities evolve as
\be \label{eq:energies_free}
    \rho_R(R)=\rhoeq\left(\frac{\Req}{R}\right)^4, \,\,\,\,\,\, \rho_\phi(R)=\rhoeq\left(\frac{\Req}{R}\right)^{\beta},
\ee
where we have defined $\beta \equiv 3(\omega+1)$ and $\rhoeq$ is the energy density of $\phi$ and radiation at $T = \Teq$, with a scale factor $R = \Req$.
Eventually, the decay of $\phi$ begins to affect the evolution of the plasma temperature, at some scale factor $R_c$. Thus, we solve analytically Eqs.~\eqref{eq:cosmo2} and~\eqref{eq:cosmo3} at first order in $\Gamma_\phi/\Heq$, where $\Heq\equiv H(\Req)$, following closely Ref.~\cite{Arias:2020qty}. We find that the energy of $\phi$ and radiation is well described by 
\begin{align}
    \rho_\phi(R) &\simeq \rhoeq\,  \left[\left(\frac{\Req}{R}\right)^\beta - \frac{2}{\beta} \frac{\Gamma_\phi}{\Heq} \left(\frac{\Req}{R}\right)^{\beta/2}\right], \label{eq:phi_density}\\
    \rho_R(R) &\simeq \rhoeq\, \left[\left(\frac{\Req}{R}\right)^4 + \frac{2}{8-\beta} \frac{\Gamma_\phi}{\Heq} \left(\frac{\Req}{R}\right)^{\beta/2}\right]. \label{eq:rad_density}
\end{align}
For $R\ll R_c$ the first terms of the rhs of Eqs.~\eqref{eq:phi_density} and~\eqref{eq:rad_density} are the dominant ones. Eventually, as the decay of $\phi$ starts to be important,  the second terms of the rhs can not be neglected. We find, by equating the first and second term of the rhs of Eq.~\eqref{eq:rad_density} and using the expressions of Eq.~(\ref{app:eq:cond}), that
\be
\Rc\simeq \Req \left( \frac{(8-\beta)}{2 } \left( \frac{\Teq}{\Tend}\right)^2 \right)^{\frac{2}{8-\beta}}.
\ee
And by requiring the second and third term of the rhs of Eq.~\eqref{eq:phi_density} are comparable, we find an expression for when the decays start to be important, meaning $R = \Rend$, to be
\be \label{eq:rend}
\Rend\simeq \Req\left(\frac{\beta }{2} \left( \frac{\Teq}{\Tend}\right)^2\right)^{2/\beta}.
\ee
where we can also find the temperature at $R = \Rc$ to be
\be
    \Tc \simeq
    \Teq\left(\frac{2}{8-\beta} \frac{\Tend^2}{\Teq^2}\right)^\frac{2}{8-\beta}.
\ee
From this analysis it can be also extracted that deep during the $\phi$ domination, the relation between temperature and scale factor is
\be \label{eq:T_NSC}
    T(R) \simeq \Teq \left[\frac{2}{8-\beta}\frac{\Tend^2}{\Teq^2} \right]^{1/4} \left(\frac{\Req}{R}\right)^{\beta/8}.
\ee
Figure~\ref{fig:energies_nsc} shows an example of the evolution of the SM and $\phi$ energy densities, for $\Teq =1$~GeV, $\Tend = 4$~MeV and $\beta = 3$ (EMD).
For the case of an EMD, or in general for $\beta < 4$, $\Req$ corresponds to the scale factor at which $\rho_\phi$ starts to dominate over $\rho_R$, $R_c$ to the scale factor where effectively $\rho_\phi$ starts to dominate the evolution of $\rho_R$, and
$\Rend$ is a proxy of the scale factor where $\phi$ decays completely.

\begin{figure}[ht!]
	\centering 
	\includegraphics[scale=0.55]{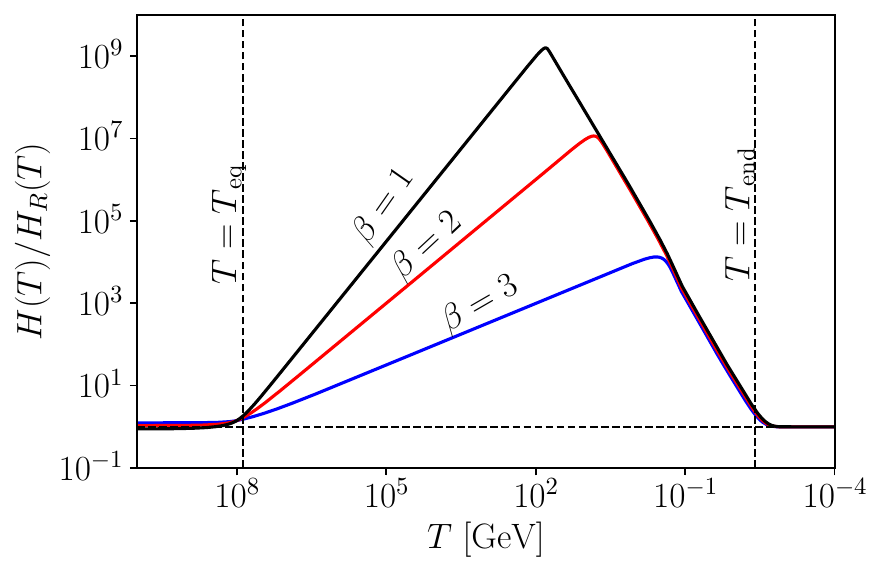}
    \includegraphics[scale=0.55]{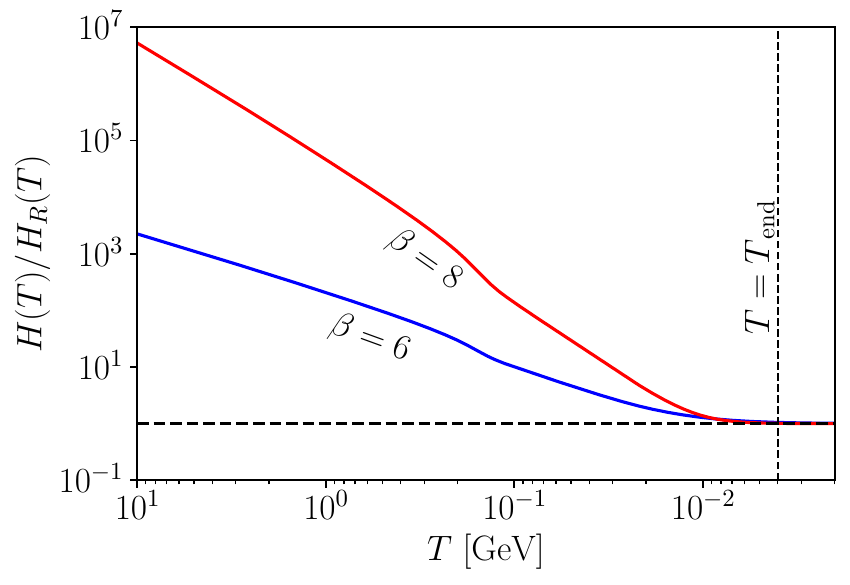}
	\caption{The evolution of the Hubble expansion rate $H$ normalised to the case dominated by radiation $H_R$, as a function of the temperature for different NSC.
	The left panel corresponds to $\Teq = 10^8$~GeV and $\Tend = 4$~MeV, for $\beta =1$, 2 and 3, whereas the right panel to $\Tend = 4$~MeV for $\beta = 6$ and 8.}
	\label{fig:H_vs_T}
\end{figure}
Taking into account the contributions from radiation and $\phi$, Fig.~\ref{fig:H_vs_T} shows the evolution of the Hubble expansion rate $H$ normalised to  case dominated by radiation $H_R$, as a function of the temperature for different NSCs.
The left panel corresponds to $\Teq = 10^8$~GeV and $\Tend = 4$~MeV, for $\beta =1$, 2 and 3, whereas the right panel to $\Tend = 4$~MeV and $\beta = 6$ and 8.
The different regions can be understood analytically.
In the case with $\beta < 4$ (left panel)
\begin{equation}
    H(T) \simeq
    \begin{dcases}
        H_R(T) &\text{ for } T \gg \Teq\,,\\
        H_R(\Teq) \left(\frac{T}{\Teq}\right)^\frac{\beta}{2} &\text{ for } \Teq \gg T \gg \Tc\,,\\
        H_R(\Tend) \left(\frac{T}{\Tend}\right)^4 &\text{ for } \Tc \gg T \gg \Tend\,,\\
        H_R(T) &\text{ for } \Tend \gg T\,,
        \label{eq:Hub2}
    \end{dcases}
\end{equation}
whereas in the opposite case with $\beta > 4$ (right panel)
\begin{equation}
    H(T) \simeq
    \begin{dcases}
        H_R(\Tend) \left(\frac{T}{\Tend}\right)^\frac{\beta}{2} &\text{ for } T \gg \Tend\,,\\
        H_R(T) &\text{ for } \Tend \gg T\,.
    \end{dcases}
\end{equation}

\section{Axion oscillation in NSC} \label{sec:nsc_oscillation}
In order to compute the relic density at present, we follow the same lines as in the standard case, 
taking into account the entropy injection due to the decay of $\phi$.
Thus, the axion energy density today gets the form
\be 
    \rho_a(T_{0}) =  \rho_a(\Tosc) \frac{\ma}{\maT(\Tosc)}\left(\frac{\Rosc}{R_0}\right)^3
    = \rho_a(\Tosc)\frac{\ma}{\maT(\Tosc)}\frac{s(T_0)}{s(\Tosc)} \times \frac{S_{\rm osc}}{S_{\rm end}}.
    \label{eq:NSC_axion_relic_density}
\ee
which is similar to Eq.~\eqref{axion_density_sc}, multiplied by a factor that takes into account the dilution of the axion energy density due to entropy injection between $\Tosc$  and $\Tend$, given by $S_{\rm osc}/S_{\rm end}$\footnote{Let us precise that in this equation, an instantaneous decay of $\phi$ has been assumed at $T=\Tend$, which of course is not true. Numerically we are taking this into account, defining a temperature/scale factor such that the field has effectively stopped to affect the SM radiation. For sake of clarity we will use $\Tend$ in our analytical expressions.}. 
The dilution factor can be expressed in terms of the parameters of the NSC, depending on which stage of the $\phi$ dominance the axion starts to oscillate. Moreover, this factor appears only for $\beta<4$, as for $\beta>4$ $\phi$ does not deposit any energy in the plasma and then is equal to one.   

In a NSC scenario, the oscillation temperature will always be smaller or equal to the oscillation temperature in a radiation dominated expansion, because $H>H_R$ (see Figs.~\ref{fig:H_vs_T} and~\ref{fig:Tosc_nsc}). Then, the effect of lowering the oscillation temperature in a NSC leads to an increased axion energy density (as will be found when analysing cosmologies with no entropy injection, {\it i.e.} $\beta>4$). On the other hand, the dilution factor in Eq.~\eqref{eq:NSC_axion_relic_density} causes the axion relic abundance to decrease by diluting its energy density.

In the rest of this section we will obtain analytical expressions for the axion relic density in terms of the parameters of the NSC and we will find the extreme range of masses that could be achieved for a given cosmology. For simplicity of the analytical expressions, we drop the dependence on the degrees of freedom ({\it{i.e.}}, they will be taken to one), as we are only interested in the order of magnitude. For the numerical results, they are fully considered, the same as the anharmonicities in the potential, as outlined in Appendix~\ref{app:anF}.
Before proceeding further, we would like to note that even if in the present study we focus on the QCD axion, this analysis could be generalised straightforwardly to axion-like particles~\cite{Blinov:2019rhb}, without the need to impose the QCD axion mass-scale relation.

\subsection[Early matter domination and $\beta < 4$]{\boldmath Early matter domination and $\beta < 4$} \label{sec:matter}

Cosmologies with $\beta<4$ correspond to equations of state with $\omega<1/3$. 
It turns useful for our detailed analysis to  divide the NSC into three stages, which will lead to different outcomes for the axion relic density, depending in which one the oscillation of the axion field takes place:
\begin{itemize}
    \item[$i)$] Region 1, $R\ll\Req$: radiation dominated universe.
    \item[$ii)$] Region 2, $\Req\ll R\ll \Rc$: dominance of $\phi$, prior its decay becomes significant.
    \item[$iii)$] Region 3, $\Rc\ll  R\ll \Rend$: dominance of $\phi$, which significantly decays into radiation, modifying the evolution of the SM temperature. After this stage, the universe returns to a radiation dominance.
\end{itemize}
\begin{figure}[t!]
    \centering
    \includegraphics[width=0.49\textwidth]{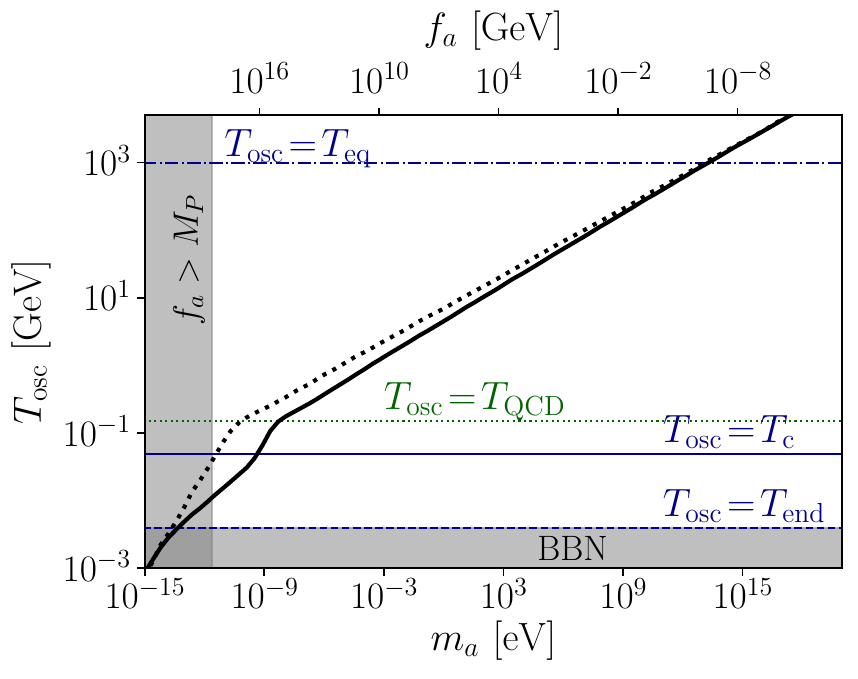}
    \includegraphics[width=0.49\textwidth]{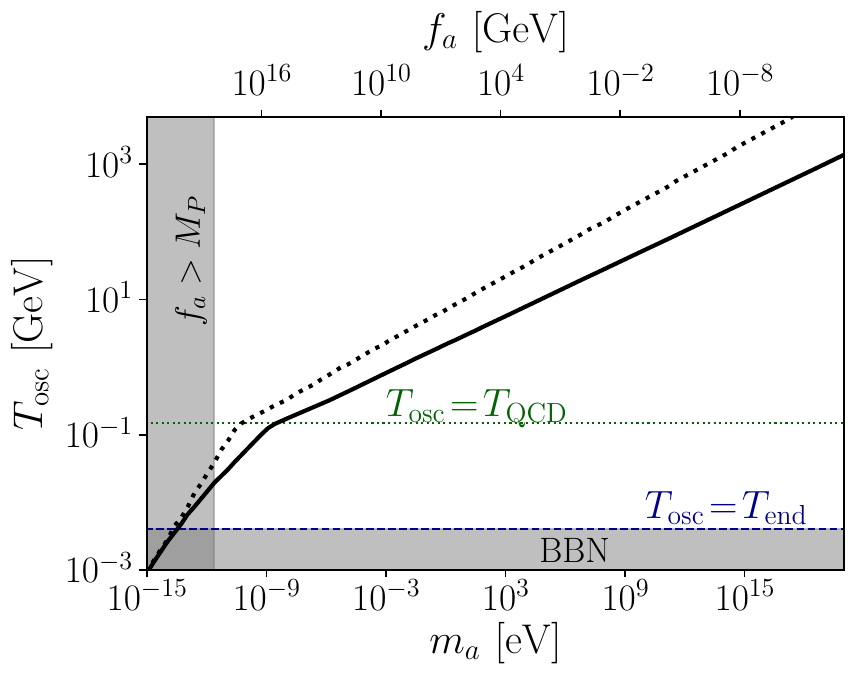}
    \caption{The oscillation temperature as a function of the axion mass for a non-standard (solid black) and standard (dotted black) cosmology. The horizontal blue lines denote the boundaries of the regions studied in section~\ref{sec:matter}. The left panel corresponds to $\beta=3$, $\Teq\simeq  10^{3}$~GeV, $\Tend= 4$~MeV, whereas the right panel to kination $\beta=6$ with $\Tend= 4$~MeV. }
    \label{fig:Tosc_nsc}
\end{figure}
In Fig.~\ref{fig:Tosc_nsc} we show the oscillation temperature $\Tosc$ vs the axion mass $\ma$ in a NSC scenario with $\beta=1$ and we have highlighted the boundaries of the three NSC regions, and can be clearly recognised that for each of them $\Tosc$ features a different dependence on $\ma$ that we will find analytically. 

\begin{figure}
	\centering
	\includegraphics[scale=0.84]{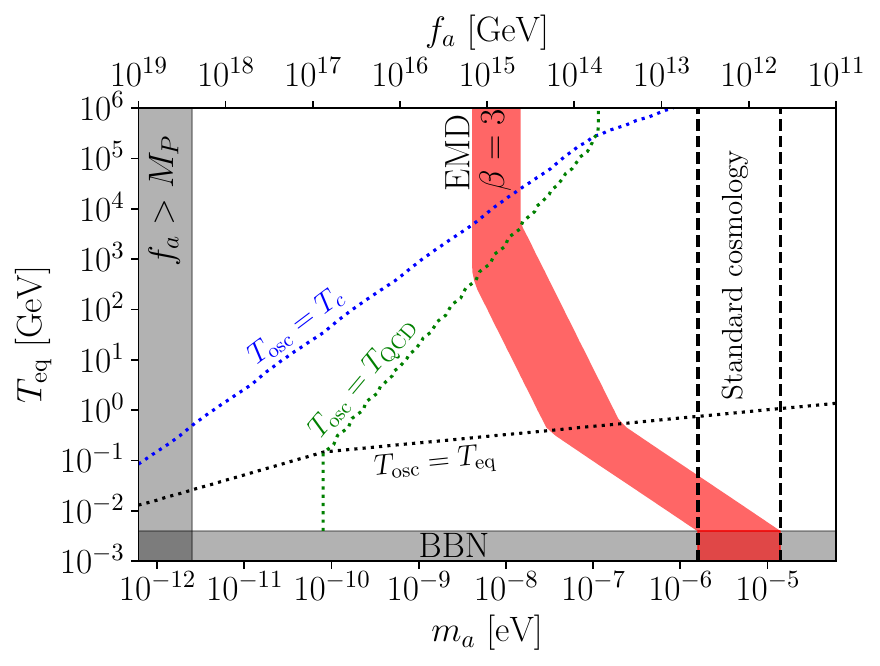}
	\includegraphics[scale=0.84]{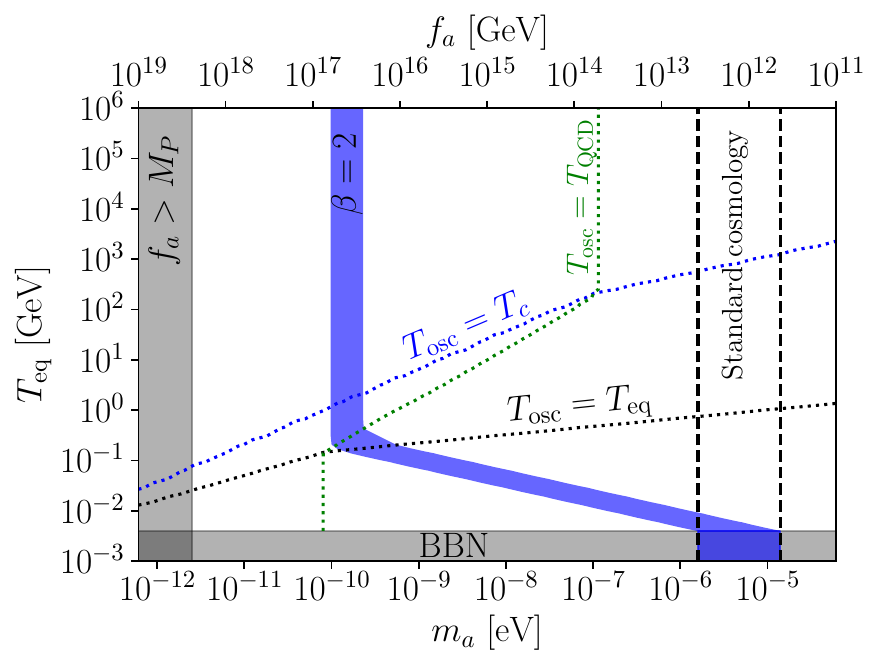}
    \caption{Parameter space corresponding to the whole observed DM abundance, for an early matter dominance case (EMD) with $\beta = 3$ (top) and $\beta = 2$ (bottom), for $\Tend = 4$~MeV and $0.5 \leq \theta_i \leq \pi/\sqrt{3}$.
    Between the vertical dashed lines the results for the standard cosmology are recovered.
    The dotted lines correspond to $\Tosc = \Teq$, $\Tosc = T_c$ and $\Tosc = \TQCD$.
    The grey bands show the regions where $\Teq < T_\text{BBN}$ or $f_a > M_P$.}
	\label{fig:matter}
\end{figure}

Our aim is to find the smallest axion mass -- equivalently, the highest possible $\fa$ scale -- that can accommodates the whole cold DM observed today for a given NSC. To do that, we start writing the entropy dilution factor for the axion in each region.
Secondly, we  find the requirements on the axion mass in order for the oscillation of the axion field to take place during any of the stages of the NSC delineated above.
The oscillation temperature can be divided into two regimes, depending on whether it takes place before or after the QCD phase transition. Therefore, we divide our analysis of each region for the case of constant axion mass (oscillations below the QCD transition) or thermal axion mass (oscillations above the QCD transition). For both regimes, we find the  smallest possible mass that gives the whole relic density, as a function of the NSC parameters, by requiring the highest possible dilution taking some NSC benchmark values. Our main results for $\beta < 4$ are summarised in Fig.~\ref{fig:matter}, where we show the parameter space that generates the whole DM abundance for $\beta = 3$ (red band, top) and $\beta = 2$ (blue band, bottom), for $\Tend = 4$~MeV. We have considered as a width a range of initial angles $\theta_i \in \left[0.5, \pi/\sqrt{3}\right]$. The regions to the left of the bands produce a DM overabundance and the ones to the right a sub-production. We compare our analytical estimates with these numerical results, finding an excellent agreement.
The region between the vertical dashed lines depicts the case of a standard cosmology.

\subsubsection[Region 1: $\Rosc \ll \Req$]{Region 1: $\boldsymbol{\Rosc \ll \Req}$}
In this region, the oscillations of the axion field occur in a period of radiation domination,  before the influence of $\phi$ and only differs from the standard cosmological scenario in that there is entropy transfer to the SM, not to axions, causing a dilution of the DM density. We can get an analytical estimation of the dilution factor, valid when the fluid gets to dominate the expansion of the universe,\footnote{Meaning valid only when the chosen parameters $\beta$, $\Teq$ and $\Tend$ lead to a solid fluid domination.} as 
\be \label{eq:gamma_R1}
    \gamma_{R_1} = \frac{S_{\rm osc}}{S_{\rm end}} \simeq \left[\left( \frac{4}{\beta^2}\right) \left(\frac{\Tend}{\Teq}\right)^{4-\beta} \right]^{3/\beta},
\ee
for a more detailed derivation, see Appendix~\ref{app:reg1}, Eq.~\eqref{app:eq:gamma_R1}. Therefore, in this regime,
\be \label{eq:axion_densityR1}
    \Omega_a = \Omega_a^{\rm std}\, \gamma_{R_1}\,,
\ee
with $\Omega_a^{\rm std}$ corresponding to the axion density in the standard cosmological scenario.
By inspecting $\gamma_{R_1}$ it can be seen that the dilution is more important for cosmologies with low $\beta$. Moreover, a higher dilution is obtained for the small $\Tend$ and large $\Teq$ values. 

The condition for the oscillation of the axion field to  occur during this period is  given by  $\Teq\ll \Tosc$. Let us analyse it for both regimes.

\subsubsection*{Constant axion mass}
The constant axion mass regime corresponds to $\Teq\ll \Tosc \lesssim \TQCD$, which can also be written in terms of the axion mass, by using the expression for the oscillation temperature in Eq.~\eqref{eq:Tosc_std}, as
\be
    4\times 10^{-12}~{\rm{eV}} \left(\frac{\Teq}{0.1~\mbox{GeV}}\right)^2 \ll \ma\lesssim 10^{-11}~\mbox{eV},
\ee
independently on $\beta$.
The axion relic abundance in this region is obtained from Eq.~\eqref{eq:axion_densityR1}. Imposing that the axion constitutes all DM of the universe, the axion mass in terms of the NSC parameters obeys
\begin{equation} 
    \ma \simeq
    \begin{dcases}
        5\times 10^{-10}\,\mbox{eV}\,\,  \left(\frac{\Tend}{\TBBN}\right)^{2} \left(\frac{0.1\, \mbox{GeV}}{\Teq}\right)^{2}\, \theta_{i}^{4/3} & \text{for } \beta=2\,,\\
        2\times 10^{-8}\,\mbox{eV}\,\,  \left(\frac{\Tend}{\TBBN}\right)^{2/3} \left(\frac{0.1\, \mbox{GeV}}{\Teq}\right)^{2/3}\, \theta_{i}^{4/3} & \text{for } \beta = 3\,,
    \end{dcases}
\end{equation}
for $\ma \lesssim 10^{-11}$~eV. In order to maximise  the impact of the  dilution, we have chosen $\Tend\sim \TBBN$.

For the two equations of state considered above, we have to satisfy that the axion mass is below $\sim 10^{-11}$eV, but that would need of a $\Teq$ which is (much) higher than $\TQCD$ or the initial misalignment angle should be taken unnaturally small. Therefore, we find that if the oscillation of the axion field happens during this regime, it can not accommodate the whole DM density and could only be a subcomponent.

\subsubsection*{Thermal axion mass}
On the other hand, for $\Tosc\gtrsim\TQCD$, the axion mass receives important thermal corrections. Then, since $\Tosc\gg \Teq$, we find
\bea
 10^{-12}\, \mbox{eV}\left(\frac{\Teq}{0.1\,\mbox{GeV}}\right)^6\ll m_a, \,\,\,\,\,\,\,\,\,\,\,\, \mbox{for}\,\,\, \ma\gtrsim 10^{-11}\rm{eV}.\label{eq:m_R1_T}
\eea
Note that $\Teq$ appears to the sixth power, so raising its value also increases the value of the axion mass that could oscillate in this period. Then, it seems difficult to open up the DM axion window much if the oscillation happens during this regime, since in order to get the highest entropy dilution, a high $\Teq$ is needed, but also this rises the axion mass range, that already in the SC subproduces the DM density.
Next, we find the mass that can accommodate the whole DM relic density
\begin{equation} \label{eq:relic_R1values}
    \ma \simeq 
    \begin{dcases}
        2\times 10^{-10}\,\mbox{eV} \left(\frac{\Tend}{\TBBN}\right)^{18/7}\left(\frac{0.2\, \mbox{GeV}}{\Teq}\right)^{18/7}\theta_i^{12/7} & \text{for } \beta = 2\,,\\
        3\times 10^{-6}\, {\rm{eV}}\,\left(\frac{\Tend}{\TBBN}\right)^{6/7} \left(\frac{0.1\, \mbox{GeV}}{\Teq}\right)^{6/7}\, \theta_{i}^{12/7} & \text{for } \beta = 3\,,
    \end{dcases}
\end{equation}
for $\ma \gtrsim 10^{-11}$~eV.
We have considered $\Tend\sim\TBBN$, to get the maximum dilution possible, thus  we also require the highest $\Teq$, but  recalling from Eq.~\eqref{eq:m_R1_T}, that increasing that temperature, makes the axion mass to rise, a compromise has to be found.
Thus, by incorporating that requirement, we find for $\beta=3$ that around $\Teq\lesssim 0.6$~GeV for $\theta_i\sim 1$, an axion of mass $\ma\sim 6\times 10^{-7}$~eV could explain the whole DM density. For $\beta=2$, it is found that the smallest axion mass that can satisfy the DM abundance is around $\ma \sim 10^{-10}$~eV, with a $\Teq\sim 0.2$~GeV. For those estimations we have assumed $\theta_i \simeq \mathcal O(1)$. These findings are supported by Fig.~\ref{fig:matter}, where a plot of the temperature $\Teq$ vs $\ma$ is shown, such that the red band corresponds to the mass range that fits the whole DM relic density observed today, with more details in the caption.  Region 2 for the case of thermal axion masses contributes to the  lower part of the plot. From that figure we confirm that in order for an axion to oscillate during the period defined by Region 2, the NSC has to be quite short, with $\TBBN \lesssim \Teq \lesssim 0.1~$GeV and $\Tend \sim \TBBN$.  Moreover, this region opens up to the range of mass just below the standard cosmological scenario, even overlapping with that cosmology, allowing to reach PQ scales as high as nearly $\fa\lesssim 10^{14}$~GeV for $\beta=3$ and $\fa\sim  10^{16}~$GeV for $\beta=2$, with initial misalignment angles $\theta_i\sim 1$.

\subsubsection[Region 2: $\Req\ll \Rosc\ll \Rc$]{Region 2: $\boldsymbol{\Req\ll \Rosc\ll \Rc}$}
In this regime, the axion oscillation happens during the domination of $\phi$, but their decays still do not affect the evolution of the temperature in the SM sector, so $T\propto R^{-1}$. 
The Hubble parameter -- assuming the $\phi$ field completely dominates the expansion -- in this case is given by
\be
    H(R) \simeq \sqrt{\frac{\rho_\phi(R)}{3\, M_P^2}}\,,
\ee
where the expression for $\rho_\phi$ to be considered is as in Eq.~\eqref{eq:energies_free}. All analytical results from this region are collected in Appendix~\ref{app:reg2}. The oscillation temperature is now dependent on the NSC parameters, so their analytical expressions can be found in Eq.~\eqref{app:eq:ToscR2}. Also the axion mass at the QCD transition it is different than in the standard cosmological scenario. For our analysis, it will be useful to write it down for some benchmark values, see Eq.~\eqref{app:eq:mR2}, 
\be \label{eq:R2_intersection_mass}
    m_{R_2} \simeq
    \begin{dcases}
        10^{-10}\,\mbox{eV}\, \left(\frac{\Teq}{2\,\mbox{GeV}}\right) &  \beta=2,\\
        2.4\times 10^{-9}\,\mbox{eV} \left(\frac{\Teq}{10^4\,\mbox{GeV}}\right)^{1/2} &  \beta=3.
    \end{dcases}
\ee

\subsubsection*{Constant axion mass}

\noindent In the case of a constant axion mass, the requirement for the oscillation to occur in Region 2 can be written as $\Tc \ll \Tosc \lesssim \TQCD$, because $\Tc < \TQCD < \Teq$ in the whole range.
Using the expression for the oscillation temperature Eq.~\eqref{app:eq:ToscR2}, we convert the oscillation temperature range into a mass range for the axion, given in  Eq.~\eqref{app:eq:NSC_R2_0bound}. Let us write it down for some benchmark values
\bea
    \beta=2; \qquad 2.7\times 10^{-11}\,\mbox{eV}\, \left(\frac{\Tend}{\TBBN}\right)^{2/3}\left(\frac{\Teq}{2~\mbox{GeV}}\right)^{4/3}\ll \ma \ll m_{R_2}, \\
   \beta=3; \qquad 9\times 10^{-10}\, \mbox{eV} \left(\frac{\Tend}{\TBBN}\right)^{6/5} \left(\frac{\Teq}{10^4~\mbox{GeV}}\right)^{4/5}\ll m_a \ll m_{R_2}. 
\label{eq:R2_mass_range_0}
\eea
Thus, for both $\beta=2$ and $\beta=3$, the mass range can lie below the one for standard cosmology, and it gets lower for $\beta=2$, but, they have  a rather small range that satisfies $\ma\lesssim m_{R_2}$, by comparing with Eq.~\eqref{eq:R2_intersection_mass}. Note that in the case $\beta=2$, we have to choose a fairly small $\Teq$ to meet the correct mass range. 

The axion relic density is computed evaluating the energy density with the corresponding  oscillation temperature in this region, and then multiplying by the dilution factor, which is
the same as Region 1, since the fluid has not yet started to decay. 
The detailed expressions for the axion relic density are computed in the appendix, Eq.~\eqref{app:eq:R2_relic}, from those expressions, we can find the axion mass that  accommodates the whole DM density in a NSC if it happens that the axion oscillates in Region 2 with constant mass. As done before, we equate the relic density in this region to $\Omega_{\rm CDM}$ and we again consider $\Tend\sim \TBBN$ to assure the maximum  dilution and the initial misalignment angle is considered $\mathcal O(1)$. We find 
\begin{equation} \label{eq:R2_relic_mass_0}
    \ma \simeq
    \begin{dcases}
        4.7 \times 10^{-11}~\text{eV} \left(\frac{\Tend}{\TBBN}\right) {\theta_i}^{2/3} & \beta = 2\,,\\
        2.7\times 10^{-9}\, \mbox{eV}   \left(\frac{\Tend}{\TBBN}\right)^{1/2}\, {\theta_i} & \beta = 3\,,
    \end{dcases}
\end{equation}
for $\ma \lesssim m_{R_2}$.
Let us first stress that the relic density is independent on $\Teq$. Secondly, we require they are   within the limits of Eq.~\eqref{eq:R2_mass_range_0}  which leads to the condition  $10^4\,\mbox{GeV}<\Teq <3.6 \times 10^4$~GeV for $\beta=3$ and $0.7\, \mbox{GeV} <\Teq<4$~GeV for $\beta=2$. Therefore, we have found that the smallest axion mass that could fit the DM relic abundance in the case of  $\beta=3$ is 
\be
    \ma\sim \mbox{few}\times 10^{-9}\,\mbox{eV}, \,\,\,\,\,\,\,\mbox{with}\,\,\,\,\,\, 10^4\, \mbox{GeV}<\Teq<3.6\times 10^4\,\mbox{GeV} ,
\ee
which can be seen from Fig.~\ref{fig:matter} that it corresponds to the region in the red band that it is roughly between the green dotted line ($\Tosc=\TQCD$) and the blue line ($\Tosc=\Tc$). Analogously, for $\beta=2$ the smallest axion mass is found roughly
\be
 \ma\sim \mbox{few}\times 10^{-11}\mbox{eV}, \quad \text{with}\quad 0.7\,\mbox{GeV}<\Teq<4\,\mbox{GeV},
\ee
which also corresponds to the region in the lower panel of Fig.~\ref{fig:matter} that lies between the dotted green line ($\Tosc=\TQCD$) and the dotted blue line ($\Tosc=\Tc$). As we shown in our analysis, this region has a short range, meaning that only for cosmologies with narrow range of temperatures $\Teq$ can give the right dilution to the relic density, such that the whole DM relic density can be produced.

\subsubsection*{Thermal axion mass}

\noindent In case the oscillation happens above the QCD  phase transition temperature, the range to be considered is $\TQCD \ll \Tosc\ll \Teq$, using expression Eq.~\eqref{app:eq:ToscR2} from the Appendix and converting the above requirement into an axion range, we get
\bea \label{eq:R2_mass_range_T}
    m_{R_2}\ll \ma\ll 4\times 10^{-8}\mbox{eV} \left(\frac{\Teq}{0.6\,\mbox{GeV}}\right)^6, 
\eea
valid for any equation of state. From the relic density found in Eq.~\eqref{app:eq:R2_relic} we obtain the axion mass that can fit the whole DM relic density to be of the order of
\begin{equation}
    \ma \simeq
    \begin{dcases}
        10^{-10}\,\mbox{eV}\left(\frac{0.1\,\rm{GeV}}{\Teq} \right)^{8/7}  \left(\frac{\Tend}{\TBBN}\right)^{15/7} \theta_i^{10/7} & \text{for } \beta = 2\,,\\
        2\times 10^{-9}\, \mbox{eV} \left(\frac{10^{4}~\rm{GeV}}{\Teq} \right)^{2/7}  \left(\frac{\Tend}{\TBBN}\right)^{11/14} \theta_i^{11/7} & \text{for } \beta = 3\,,
    \end{dcases}
\end{equation}
for $\ma\gtrsim m_{R_2}$.
Requiring they belong to the mass range in Eq.~\eqref{eq:R2_mass_range_T}, we get that the equilibrium temperature has to satisfy $0.6\,\mbox{GeV}< \Teq < 10^4$~GeV for $\beta=3$ and $0.2\, \mbox{GeV}< \Teq< 0.7\, \mbox{GeV}$ for $\beta=2$. Comparing with our numerical result from Fig.~\ref{fig:matter} we find a good agreement. In the case of EMD ($\beta=3$) there is a much broader range of cosmologies that can lead to the right amount of DM production. But, because the relic has a steeper  dependence on the mass than in the standard cosmological scenario ($\ma^{-14/11}$ vs. $\ma^{-7/6}$), the range of masses covered is not as broad as in the previous region. Instead, for $\beta=2$ our analysis supported by the lower panel of Fig.~\ref{fig:matter} shows that the parameter space of NSC such that the axion DM is produced during this regime it is extremely narrow (between black and green dotted lines) and the axion mass is set to be between $10^{-10}-10^{-9}$ eV, which actually opens up the PQ scale more than for a EMD era, to near $\fa \sim 10^{17}$ GeV.

\subsubsection[Region 3: $\Rc\ll \Rosc\ll \Rend$]{Region 3: $\boldsymbol {\Rc\ll \Rosc\ll \Rend}$}
For this regime, oscillations of the axion field happen during the domination and decay of  $\phi$.  The Hubble parameter is taken to be $H\simeq \sqrt{\frac{\rho_\phi}{3 M_P}}$ and temperature and scale factor are related by $T\propto R^{-\beta/8}$, see Eq.~\eqref{eq:T_NSC}. Therefore, this time results more convenient to impose the condition $\Rc\ll \Rosc\ll \Rend$, than using the temperatures. Full analytical expressions used for this region can be found in Appendix \ref{app:reg3}. 
Let us also remind that during this period, due to the entropy injection, the expansion of the universe follows the behaviour $H\propto T^4$. The dilution to the axion energy  is different than in the previous regions, because $\phi$ is already decaying. Using $\Rend$ given in Eq.~(\ref{eq:rend}) we find
\be
    \gamma_{R_3} = \frac{S_{\rm osc}}{S_{\rm end}} = \left(\frac{\Tosc}{\Tend} \frac{R_{\text{osc}}}{\Rend}\right)^3 \simeq \left(\frac{\Tend}{\Tosc}\right)^{24/\beta-3}.
\ee
For smaller $\beta$, as expected, the dilution is more important. On the other hand, since $\Tend/\Tosc$ is bigger than the ratio $\Tend/\Teq$ from Eq.~\eqref{eq:gamma_R1}, we obtain the relation $\gamma_3 > \gamma_1$, since in Region 3 the field $\phi$ is already decaying, the entropy injection into the thermal bath is smaller than in the previously considered regions. The highest dilution of the axion energy density will be attained for the smallest $\Tend$, so we shall  again choose $\Tend = \TBBN$, for our analysis.

For the oscillation temperatures, the expressions have a very mild dependence on $\beta$, that does not change the order of magnitude of the expressions (see Eq.~\eqref{app:eq:ToscR3}), so we write them as valid for any equation of state
\begin{align} \label{eq:R3_Tosc}
    \Tosc^{R_3} \simeq
    \begin{dcases}
        0.04~\mbox{GeV} \left(\frac{\ma}{10^{-4}\,\mu\mbox{eV}}\right)^{1/4} \left(\frac{\Tend}{10~\mbox{MeV}}\right)^{1/2} &  \ma\lesssim m_{R_3}\,,\\
        1~\mbox{GeV} \left(\frac{\ma}{1\,\mu \mbox{eV}} \right)^{1/8} \left( \frac{\Tend} {10~\mbox{GeV}}\right)^{1/4} &\ma\gtrsim m_{R_3}\,,
    \end{dcases}
\end{align} 
where  $m_{R_3}$ is the mass at the QCD phase transition, given by Eq.~(\ref{app:eq:mass_R3})
\be \label{eq:mR3}
    m_{R_3} \simeq 3\times 10^{-8}~\rm{eV} \left(\frac{\TBBN}{\Tend}\right)^2.
\ee
Let us make a few comments about these expressions. Firstly, there is no dependence of $\Teq$. Secondly, since we have considered  $\Tend\sim \TBBN$, the mass in Eq.~\eqref{eq:mR3} is the highest at which the transition could happen.
Recalling that in the SC the transition mass is $\sim 10^{-11}$~eV,  the QCD phase transition is shifted to a slightly higher mass in this scenario.

\subsubsection*{Constant axion mass}
\noindent The mass range for the oscillation to happen during Region 3 can be obtained from  $\Rc\ll \Rosc\ll \Rend$.\footnote{In this case we do not include the scale factor at the QCD phase transition temperature, $R_{\rm QCD}$, because it is found is always smaller than $\Rosc$ for both, constant and thermal mass regimes.} We relate the scale factor to the temperature using Eq.~\eqref{eq:T_NSC} and using the oscillation temperature found above,  we get
\bea 
    \beta&=&2; \quad    7\times 10^{-15}\, \mbox{eV}\left(\frac{\Tend}{\TBBN}\right)^2\ll \ma \ll 5\times 10^{-11}\, \mbox{eV} \left(\frac{\Tend}{\TBBN}\right)^{2/3} \left(\frac{\Teq}{4\, \mbox{GeV}}\right)^{4/3}, \label{eq:R3_mass_range_2}\\
    \beta&=&3; \quad 4\times 10^{-15}\, \mbox{eV}\left(\frac{\Tend}{\TBBN}\right)^2\ll \ma \ll  3\times 10^{-9}\, \mbox{eV}\left(\frac{\Tend}{\TBBN}\right)^{6/5} \left(\frac{\Teq}{10^5\,\mbox{GeV}}\right)^{4/5},  \label{eq:R3_mass_range_3}
\eea
for $\ma\lesssim m_{R_3}$. The mass range is in principle quite broad and it moves to higher masses by increasing $\Tend$ and only the upper bound depends on $\Teq$. All the masses from the range are restricted to be smaller than the mass at the QCD temperature, Eq.~\eqref{eq:mR3}, which limits $\Teq< 10^6$~GeV for $\beta=3$ and $\Teq<12~$GeV for $\beta=2$.

The analytical expression that accounts for the relic density during this  period is given in Eq.~\eqref{app:eq:R3_relic}. From there, we find the axion masses that accounts for the whole DM density for our two benchmark cosmologies, $\beta=2$ and $\beta=3$ are
\bea
    \ma\simeq 
    \begin{dcases}
        5\times 10^{-11}\, \mbox{eV}\,\left(\frac{\Tend}{\TBBN}\right)\, \theta_i^{2/3}, & \qquad\beta=2,\\
        2.7\times 10^{-9}\, \mbox{eV}\,\left(\frac{\Tend}{\TBBN}\right)^{1/2}\, \theta_i, &\qquad\beta=3.
    \end{dcases}
\eea
And requiring they are consistent with the mass ranges found in Eqs.~\eqref{eq:R3_mass_range_2} and~\eqref{eq:R3_mass_range_3}, respectively, it is found that $\Teq>4$~GeV for $\beta=2$ and $\Teq>8\times 10^4~$GeV for $\beta=3$. These estimates are supported by Fig.~\ref{fig:matter}, where region 3  for constant axion mass starts above the blue line $\Tosc=\Tc$. There, as found in our analysis, the relic density is independent of $\Teq$ and the smallest axion mass that can fulfil the whole DM density is, for $\beta=3$
\be
    \ma \sim  10^{-8}\, \mbox{eV}\qquad \mbox{for} \quad \Teq \gtrsim 10^4\, \mbox{GeV}, 
\ee
and for $\beta=2$
\be
    \ma \sim  10^{-10}\, \mbox{eV}\qquad \mbox{for} \quad \Teq \gtrsim 4\, \mbox{GeV}. 
\ee
Numerically in Fig.~\ref{fig:matter}, we have found that Region 3 starts to happen for constant mass axions above the blue line $\Tosc=\Tc$, in good agreement with our analysis and the numbers estimated above. Thus, for approximately $\Teq\gtrsim 10^3$~GeV, the relic density is independent on $\Teq$, where the first part corresponds to the oscillation of the axion prior to the decay of $\phi$ (Region 2) and for higher $\Teq$ when the decays are already happening, in a mass range of few $\times 10^{-8}$~eV in mass and below, for a NSC with $\Tend\sim \TBBN$ and $\theta_i\sim 1$.

\subsubsection*{Thermal mass}
\noindent If the oscillation of the axion field happens before the QCD phase transition, the range to be considered is $\Rc\ll \Rosc\ll \Rend$, and it is mapped into the mass range by
\bea
    \beta&=&2; \qquad m_{R_3}\ll \ma\ll 10^{-7}\, \mbox{eV} \left(\frac{\Tend}{\TBBN}\right)^{10/3}\left(\frac{\Teq}{10^3\, \mbox{GeV}}\right)^{8/3}, \label{eq:R3_range_b2}\\
    \beta&=&3;\qquad m_{R_3}\ll \ma\ll 5\times 10^{-7}\, \mbox{eV} \left(\frac{\Tend}{\TBBN}\right)^{22/5}\left(\frac{\Teq}{10^7\, \mbox{GeV}}\right)^{8/5}. \label{eq:R3_range_b3}
\eea
Now using the equation for the relic density in this region, given by Eq.~\eqref{app:eq:R3_relic} we obtain the axion mass that accounts for the whole DM relic density to be approximately
\bea \label{eq:R3_relic_mass}
    \ma\simeq
    \begin{dcases}
        2\times 10^{-11}\, \mbox{eV} \left(\frac{\Tend}{\TBBN}\right)^{5/2} \theta_i  & \qquad \beta=2\,,\\
        10^{-8}\, \mbox{eV} \left(\frac{\Tend}{\TBBN}\right)^{4/3} \theta_i^{4/3} &\qquad \beta=3\,.
    \end{dcases}
\eea
For $\ma \gtrsim m_{R_3}$.These expressions again do not depend on $\Teq$ so if $\Tend\sim \TBBN$, they correspond to the lowest possible mass that can account for the whole DM relic density. Comparing with the intersection mass given in Eq.~\eqref{eq:mR3}, they are slightly higher, meaning the relic density can not be completely fulfilled when axions oscillate before the QCD phase transition for the range of masses found in Eqs.~\eqref{eq:R3_range_b2} and~\eqref{eq:R3_range_b3}. We conclude that axions in those mass ranges can only account for the whole DM density if higher values of $\Tend$ are chosen,  $\Tend>1.4~\TBBN$ for $\beta=3$ and $\Tend>5~\TBBN$ for $\beta=2$.

\subsection[Kination and $\beta > 4$]{\boldmath Kination and $\beta > 4$} \label{sec:kination}
In this case, the equation-of-state parameter $\omega > 1/3$ (and hence $\beta > 4$), implying that the energy density of $\phi$ dilutes faster than the one of radiation.
Therefore, even if $\phi$ does not decay, it will naturally become subdominant, giving rise to the onset of a radiation-dominated era, at a temperature $T = \Tend$ defined by the equality $\rho_\phi(\Tend) = \rho_R(\Tend)$.
It is interesting to note that as $\phi$ does {\it not} decay, the SM entropy is always conserved.
A typical example of $\omega > 1/3$ corresponds to kination~\cite{Caldwell:1997ii, Sahni:1999gb}, where $\omega = 1$ ({\it i.e.} $\beta = 6$), which implies that $H(T) \propto T^3$, or to the case where $\omega = 5/3$ ({\it i.e.} $\beta = 8$)~\cite{Khoury:2001wf, DEramo:2017gpl} for which $H(T) \propto T^4$.

Additionally, the oscillation temperature takes the analytical expression
\begin{equation}
    \frac{\Tosc}{\Tend} \simeq
    \begin{dcases}
        \left[\frac{\gss(\Tend)}{\gss(\Tosc)}\right]^{1/3}
        \left[\frac{\ma}{3\, H_R(\Tend)}\right]^{2/\beta} &\text{ for } \Tosc \leq \TQCD\,, \\
       \left[\frac{\gss(\Tend)}{\gss(\Tosc)}\right]^\frac{\beta}{24+3\beta} 
       \left[\frac{\ma}{3\, H_R(\Tend)} \left(\frac{\Tend}{\TQCD}\right)^{-4}\right]^\frac{2}{8 + \beta} &\text{ for } \Tosc \geq \TQCD\,,
    \end{dcases}
\end{equation}
for $\Tosc \gg \Tend$.
In the opposite case where $\Tosc \ll \Tend$, the oscillation takes place during the standard cosmology, and therefore the NSC has no effect on the axion dynamics.

Note that the oscillation temperature for this scenario is in correspondence with the one  found for $\beta<4$ in region 2, with the replacement $\Teq\rightarrow \Tend$. This comes as no surprise, since cosmologies with $\beta>4$ are characterised by the dominance of $\phi$, without decaying, up to the temperature $\Tend$. The same thing happens for region 2, but the equality temperature $\Teq$ has to be identified with $\Tend$. Then, the mass at the QCD transition is given by the same expression as the one found for region 2 in Eq.~\eqref{app:eq:mR2} with the replacement $\Teq\rightarrow \Tend$, which is
\be \label{eq:kin_intersection_mass}
    m_{\rm{QCD}} \simeq
    \begin{dcases}
         4\times 10^{-10}\,\mbox{eV} \left(\frac{\TBBN}{\Tend}\right) & \mbox{for } \beta=6\,,\\
          10^{-8}\,\mbox{eV}\, \left(\frac{\TBBN}{\Tend}\right)^2 & \mbox{for } \beta=8\,.\\
    \end{dcases}
\ee
To find the axion mass range for which the oscillation is possible during this period, we only have to impose $\Tend\ll \Tosc$, using the  appropriate expression for $\Tosc$, depending if the scale is higher or lower than the QCD temperature, $\TQCD$.
\subsubsection*{Constant mass}
In the case of constant axion mass, {\it i.e.} $\Tosc\lesssim \TQCD$, we find the allowed mass range for oscillation in this regime to be
\bea
    \qquad 7\times 10^{-15}\, \mbox{eV} \left(\frac{\Tend}{\TBBN}\right)^2\ll \ma \lesssim m_{\rm QCD},
\eea
valid for any $\beta$. Secondly, we again can borrow the expression found in the analysis of region 2 for the relic density Eq.~\eqref{app:eq:R2_relic}, with the dilution factor of regions 1 and 2, Eq.~\eqref{eq:gamma_R1} equal to one. We find the expression
\bea
\Omega_a=
   \Omegastdo \left(\frac{\Tend^2}{\ma M_p}\right)^{\frac{12-3\beta}{2\beta}},\qquad \ma \lesssim m_{\rm QCD},
\eea
where $\Omegastdo$ is the relic density in the standard cosmological scenario for the constant mass regime, defined in Eq.~\eqref{app:eq:Omega0STD}.  From there, it is straightforward to find the axion mass that can account for the whole DM relic density, imposing $\Omega_a=\Omega_{\rm CDM}$ for our benchmark values $\beta=6$ and $\beta=8$
\bea
    \ma\simeq 
    \begin{dcases}
        2.4\times 10^{-3}\, \mbox{eV} \left(\frac{\TBBN}{\Tend}\right) \theta_i^2, & \qquad \beta=6\,,\\
        17\, \mbox{eV} \left(\frac{\TBBN}{\Tend}\right)\theta_i^{8/3}, & \qquad \beta=8\,,
    \end{dcases}
\eea
valid for $\ma\lesssim m_{\rm QCD}$. We can immediately note that they are significantly higher than the requirement on the mass for both cosmologies. Increasing the value of $\Tend$ does not help, because this also increases the mass at the QCD phase transition Eq.~\eqref{eq:kin_intersection_mass}. Thus, it is concluded that an axion of constant mass can not account for the full DM relic density for these cosmologies.

\subsubsection*{Thermal mass}
For the range of temperatures $\Tosc\gtrsim \TQCD$, has to be required on the one hand that $\Tosc\gg \Tend$, which leads to a requirement on $\Tend$ 
\be \label{eq:kin_Tend}
    \Tend \ll1.5\, \mbox{GeV}\, \left(\frac{\ma}{10^{-5}\, \mbox{eV}}\right)^{-1/6}.
\ee
On the other hand, the axion mass has to be above the mass at the QCD transition, thus $\ma\gtrsim m_{\rm QCD}$.
To find an expression for the relic density, we can again use the result found for region 2, Eq.~(\ref{app:eq:R2_relic}), with the dilution factor reduced to one, which gives
\be \label{eq:kination_relic}
    \Omega_a = \OmegastdT \left[\frac{\left(\TQCD^4\, \ma\, M_P\right)^{1/6}}{\Tend}\right]^{7 \frac{\beta-4}{\beta+8}}.
\ee
With $\OmegastdT$ the axion relic density in the standard cosmology for the case of thermal axion mass, defined in Eq.~\eqref{app:eq:OmegaTSTD}.
Then, the mass required to produce the whole DM relic density is given by
\bea
    \ma\simeq
    \begin{dcases}
        2.4 \times 10^{-3}\, \mbox{eV} \left(\frac{\TBBN}{\Tend}\right)\,\theta_i^2 & \qquad \beta=6,\\
        0.9\, \mbox{eV} \left(\frac{\TBBN}{\Tend}\right)^2\,\theta_i^{16/7} & \qquad \beta=8.
    \end{dcases}
\eea
Let us remark that these masses -- especially in the case of $\beta=8$ -- are higher than the standard axion DM window. In the absence of dilution, a lower oscillation temperature compared to the one in the standard scenario can be reached, increasing the axion relic density, and therefore allowing to explore higher axion masses. On the other hand, for higher values of $\Tend$ it is possible to reach smaller  masses, but recalling the requirement imposed by Eq.~\eqref{eq:kin_Tend}, we find that the maximum $\Tend$ temperature compatible with the axion as the DM solution is
\be \label{eq:kin_tend_range}
    \TBBN\lesssim \Tend\lesssim 1.4~\mbox{GeV},
\ee
valid for both $\beta=6$ and $\beta=8$ and for $\theta=1$.

\begin{figure}[t!]
	\centering
	\includegraphics[scale=1]{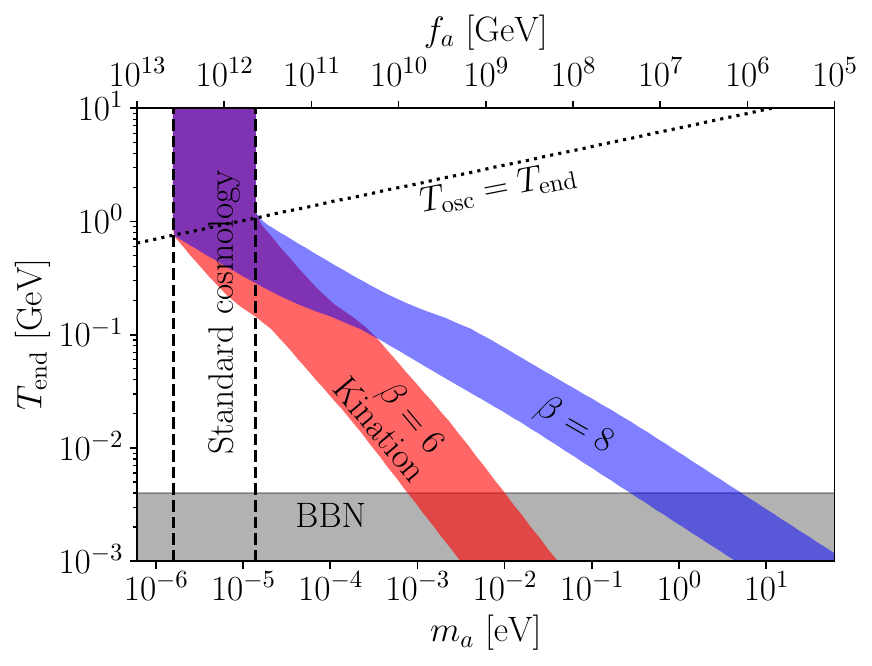}
    \caption{Parameter space corresponding to the whole observed DM abundance, for kination $\beta = 6$ (red) and $\beta = 8$ (blue), for $0.5 \leq \theta_i \leq \pi/\sqrt{3}$.
    Between the vertical dashed lines the results for the standard cosmology are recovered.
    The dotted line corresponds to $\Tosc = \Tend$, whereas in the grey band $\Tend < \TBBN$.}
	\label{fig:kination}
\end{figure} 
Figure~\ref{fig:kination} shows the parameter space generating the whole observed DM abundance, for kination $\beta = 6$ (red) and $\beta = 8$ (blue).
The widths of the bands correspond to $0.5 \leq \theta_i \leq \pi/\sqrt{3}$.
Above the dotted black line $\Tosc < \Tend$, which implies that the axion is produced in a standard cosmology, and therefore is independent on $\omega$ and $\Tend$.
However, below the dotted line (i.e., for $\Tosc > \Tend$) bigger values for the axion mass could be explored, especially for large values of $\beta$ and small $\Tend$.
The white region in the left of the bands generates a DM overabundance which overcloses the universe, whereas the region in the right can only account for a fraction of the total DM relic density.
In agreement with our analytical estimations, from Fig.~\ref{fig:kination} it can be seen that both cosmologies are spawn approximately over the region for $\Tend$ found in Eq.~\eqref{eq:kin_tend_range}. Note that the whole parameter space can be explained from the analysis made for the thermal mass scenario, $\Tosc\gtrsim \TQCD$. The apparent change in slope of both bands around $\Tend\sim 0.1$~GeV is due to the change in the relativistic degrees of freedom (ignored in our analytical treatment).  A cosmology with $\beta=8$ can accommodate the DM relic density over a higher range of masses, because  the dependence on the mass is softer than for kination and  in the standard scenario, as can be found from the expression found for the relic density, Eq.~\eqref{eq:kination_relic}. In particular, for kination ($\beta = 6$) the classical window is shifted to $1.6 \times 10^{-6}$~eV~$\lesssim \ma \lesssim 10^{-2}$~eV, whereas for $\beta = 8$ one can reach $1.6 \times 10^{-6}$~eV~$\lesssim \ma \lesssim 6$~eV.
Such higher values can be explored in the near future by experiments like MADMAX~\cite{Caldwell:2016dcw, MADMAX:2019pub, Egge:2020hyo, Beurthey:2020yuq} and (baby)IAXO~\cite{Armengaud:2014gea, Dafni:2018tvj, IAXO:2019mpb}. 
 
\section{Axion coupling to two photons}\label{sec:axion_coupling}
\begin{figure}
	\centering 
	\includegraphics[scale=0.43]{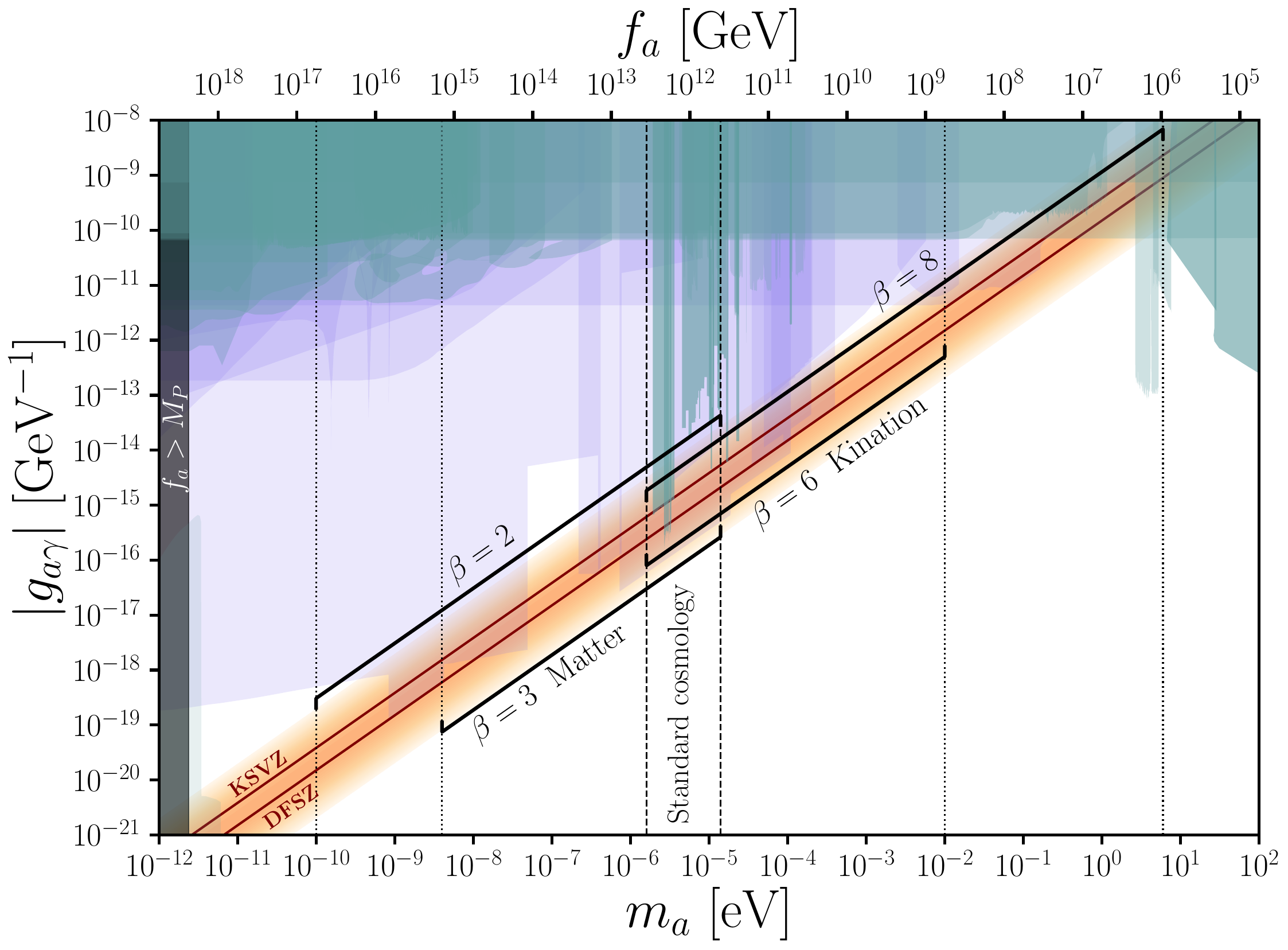}
    \caption{The axion parameter space for the axion-photon coupling in the SM and in NSC scenarios considered here, with  benchmark scenarios: $\beta=2$ ($\omega=-1/3$), $\beta=3$ (EMD, $\omega=0$), $\beta=6$ (kination, $\omega=1$) and $\beta=8$ ($\omega=5/3$). The green regions correspond to current cosmological, astrophysical and laboratory constraints. The light purple area are projections for experimental searches, as described in the text. For the standard cosmology window of axion mass we have assumed contributions to the relic abundance from the misalignment mechanism only, with the initial angle on the  range $\theta_i\in \left[0.5, \pi/\sqrt{3}\right]$, in order to include both, the pre- and post-inflationary scenarios. Figure adapted from~\cite{ciaran_o_hare_2020_3932430}.}
    \label{fig:ga_NSC}
\end{figure}

The coupling of axions to two photons is one of the most exploited to look for  signatures in observations and experimental searches. The interaction takes the form
\be
    \mathcal L_{a\gamma} = -\frac{1}4\, g_{a\gamma}\, a\, F_{\mu\nu} \tilde F^{\mu\nu} = g_{a\gamma}\, a\, \vec E\cdot \vec B\,,
\ee
where the coupling constant $g_{a\gamma}$ is model dependent and is related to the PQ scale $f_a$ as  
\be 
    g_{a\gamma}=\frac{\alpha}{2\pi f_a}\left(\frac{E}{N}-\frac{2}3\frac{4+z}{1+z} \right)\simeq 10^{-13}\, \mbox{GeV}^{-1} \left(\frac{10^{10}\mbox{GeV}}{f_a}\right), 
\ee 
where $z \equiv m_u/m_d$ and $E$ and $N$ are the electromagnetic and colour anomalies associated with the axion anomaly. For KSVZ models $E/N=0$~\cite{ kim1979weak,  shifman1980can }, whereas for DFSZ models $E/N=8/3$~\cite{Zhitnitsky:1980tq, dine1981simple}. Our findings can be mapped into the $(g_{a\gamma}, \ma)$ to see in a comprehensive way the impact of a NSC period over the axion relic density and the consequences for the axion search. Figure~\ref{fig:ga_NSC} shows the parameter space of QCD axions (yellow band) and the exclusion bounds from several astrophysical observations, cosmology based constraints and laboratory searches (brownish coloured area). The axion DM relic density generated in a  standard cosmology is bounded between the dashed lines. That is, the relic abundance obtained from the misalignment mechanism for initial angles in the range $\theta_i \in \left[0.5, \pi/\sqrt{3}\right]$, such that  both, pre-inflationary and post-inflationary scenarios are included without invoking a fine-tuned solution. The light blue area corresponds to the experimental concepts with their corresponding projections of sensitivity, all with particular emphasis to cover the QCD axion band.
We have delineated with brackets the range of masses -- equivalently, the range in the scale $\fa$ --  that a NSC with the given equations of state could access. Our benchmark values are $\beta=2$, equivalent to $\omega=-1/3$, $\beta=3$, corresponding to an EMD, {\it{ i.e.}} $\omega=0$,  $\beta=6$, known as a kination regime, with $\omega=1$ and finally $\beta=8$, which corresponds to an equation of state with $\omega=5/3$. We have fixed $\Tend = \TBBN = 4$~MeV and we scan in the equality temperature $\Teq$ according to the results obtained in subsection~\ref{sec:matter} for $\beta<4$ and the scan is in $\Tend$ for $\beta>4$, as shown in subsection~\ref{sec:kination}, all summarised in  Figs.~\ref{fig:matter} and~\ref{fig:kination}. There, it was found that for a EMD cosmology, the same mass range as the standard axion CDM window can be reached -- for cosmologies with very low $\Teq$ -- and then down to $\ma \gtrsim 4\times 10^{-9}$~eV, for cosmologies with  $\Teq\gtrsim 10^3~$~GeV. The latter is translated into a $g_{a\gamma} \sim 10^{-19}$~GeV$^{-1}$. Meanwhile, for cosmologies with $\beta=2$ it is possible to get into even smaller masses, thanks to the higher dilution of the axion relic density. According to our findings, the smallest mass attainable for this cosmology is of $\ma \gtrsim 10^{-10}~$eV, for $\Teq\gtrsim 0.1$~GeV, which corresponds to a coupling of the order of $g_{a\gamma}\sim 10^{-20}$~GeV$^{-1}$. Both scenarios will  be in the reach of the ABRACADABRA, KLASH and the next generation of ADMX.

On the other hand, cosmologies with $\beta>4$ open up the axion DM window to the right of the standard cosmological scenario, because there is no dilution of the energy density, but a smaller $\Tosc$, delaying the onset of DM production, increasing the relic density. These cosmologies will be on the reach of experiments such as ADMX, CULTASK, MADMAX, and ORGAN. For $\beta=6$ we have found the highest axion mass that can accommodate the whole DM relic density is $\ma \lesssim 10^{-2}$~eV, with a temperature $\Tend$ very close to the BBN value. The corresponding scale for the coupling to photons is $g_{a\gamma}\sim 10^{-13}$~GeV$^{-1}$. For $\beta=8$ and the same scale on $\Tend$, it is even possible to reach  $\ma \lesssim 6$~eV, which gives $g_{a\gamma}\sim 10^{-9}$~GeV$^{-1}$. Cosmologies with $\beta=8$ will be already probed with the future helioscope experiment IAXO. 
We also note that cosmologies with $\beta=8$ and $\Tend \lesssim 10^{-2}$~GeV conflict with the horizontal branch bound and other astrophysical observations~\cite{Regis:2020fhw, Grin:2006aw, Arias:2012az}. Let us note that the SN 1987A bound on the QCD axion rules out the parameter space 15~meV $\lesssim \ma \lesssim 1$~keV~\cite{Raffelt:1990yz, Carenza:2019pxu}. However,  in Ref.~\cite{Carenza:2020cis} has been pointed out that pion-related  processes  have been previously underestimated and  a better quantitative understanding of their contribution is needed. Thus, it still seems worth to explore this region of parameter space independently with experimentally well-known and controlled sources.
For an exhaustive review on the experimental programme of direct detection searches and expected timescales, we refer to Refs.~\cite{Irastorza:2018dyq, Sikivie:2020zpn, Billard:2021uyg}.
Finally, let us remark that for all the equations of state studied here, it is possible to partially cover the standard axion DM mass range, although this happens for a very short-lived NSC.

\section{Conclusions} \label{sec:conclusions}
Throughout this paper we have made a comprehensive and exhaustive study of the implications of a nonstandard cosmological (NSC) period -- somewhere prior to the BBN era -- on the onset of the axion field oscillation and the corresponding dark matter (DM) relic density generated in the so-called process of misalignment. We have considered a generically new field $\phi$ whose energy content gets to dominate the expansion of the universe, until it decays into SM radiation for $\omega<1/3$ or dilutes away in the case of $\omega>1/3$, with $\omega$ the equation-of-state parameter. As examples, we have considered benchmark cosmological scenarios that have been motivated previously in the literature: $\omega=-1/3$, $\omega=0$ (early matter domination), $\omega=1$ (kination) and $\omega=5/3$. Although we have also found analytical expressions for a generic equation of state.

In the case of $\omega<1/3$, we have divided the period of NSC in three regions: {\it{i)}} a radiation driven period before the dominance of $\phi$, {\it{ii)}} domination of $\phi$ prior to its decay and {\it{iii)}} the era of decay and corresponding entropy injection into the SM plasma. For all these regions, we have found the corresponding oscillation temperature, DM relic density and the axion mass as a function of the NSC parameters that can account for the whole DM density observed today, distinguishing the regime of temperatures above the QCD phase transition (where the mass gets important thermal effects) and below (where it can be considered constant).  Our findings have shown that in principle, a wide range of cosmologies, with their characteristic parameters: $\omega$, $\Teq$ (temperature that sets the beginning of the dominance of $\phi$ over radiation) and $\Tend$ (temperature where $\phi$ has mostly decayed away) can lead to the right production of axion DM in a wide range of masses below the standard axion window. For the case of $\omega=-1/3$, we have found that the smallest possible axion mass (without a tuning of the initial misalignment angle) is around $\ma \sim 10^{-10}~$eV and can be achieved for cosmologies with $\Teq\gtrsim 0.1$~GeV, $\Tend\sim 4$~MeV. In that case, the oscillation of the axion field happens  either before or during the epoch of decay of $\phi$ and below the QCD phase transition, so the axion mass can be considered constant. Something similar is found for a NSC with matter domination $\omega=0$, where the smallest axion mass is around $\ma \sim 10^{-8}$~eV, for cosmologies with $\Teq \gtrsim 10^2$~GeV, $\Tend\sim 4$~MeV, as shown in Fig.~\ref{fig:matter}.

For cosmologies with $\omega>1/3$, we have considered a stable $\phi$, with an energy density that fades away due to the expansion of the universe. In that case there is no dilution, and we find that the axion relic density is higher than in the standard cosmological scenario, due to a lower oscillation temperature -- consequence of a bigger Hubble parameter -- than in the standard radiation-dominated scenario. Thus, higher axion masses than in a radiation-dominated universe can lead to the explanation of the DM paradigm. For the equations of state $\omega=1$ and $\omega=5/3$,  we have found that cosmologies where the extinction of $\phi$ takes place near the BBN epoch can lead to higher axion masses, $\ma \sim 10^{-2}$~eV and $6$~eV, respectively. In this case, the axion oscillation happens above the QCD phase transition and therefore the mass receives thermal corrections, as shown in Fig.~\ref{fig:kination}. 

Concerning the impact of NSC on axion searches, we have projected our results into the exclusion plot of the axion coupling to two photons, the most exploited interaction to search for these particles. We have highlighted that a handful of NSCs can spawn the axion DM window in the range $10^{-20}\, \mbox{GeV} \lesssim g_{a\gamma}$,  all the way up to $g_{a\gamma}\sim 10^{-9}$~GeV, see Fig.~\ref{fig:ga_NSC}, the upper part being in slight conflict with astrophysical searches. Remarkably, most of the parameter space is expected to be probed by experimental prospects in the near future. These findings are expected to have a positive impact on the motivation to search with laboratory-based experiments for axion DM outside the so-called classical window of standard cosmology. Then, a potential discovery of the QCD axion will shed some light into the cosmological history of the early universe, prior to the BBN era.

Let us finally stress that we have only considered the production of DM through the misalignment mechanism. In the case where the PQS is broken after the inflationary epoch, the potential contribution from topological defects such as strings and domain walls could move the mass ranges found here. Nonetheless, given the still ongoing uncertainty on the fraction contributed by these mechanisms, we have only focused on the misalignment production.

\section*{Acknowledgments}
We are thankful to Luca Visinelli for useful comments and feedback. PA and MV acknowledge support from FONDECYT project 1161150 and Proyecto POSTDOC-DICYT, 042131AR-AYUDANTE, VRIDEI. PA acknowledges support from DICYT 042131AR project. NB received funding from the Spanish FEDER/MCIU-AEI under grant FPA2017-84543-P, and the Patrimonio Autónomo - Fondo Nacional de Financiamiento para la Ciencia, la Tecnología y la Innovación Francisco José de Caldas (MinCiencias - Colombia) grant 80740-465-2020.
DK and LR are supported in part by the National Science Centre, Poland, research grant No.  2015/18/A/ST2/00748.   LR is also supported  by  the  project AstroCeNT:  Particle  Astrophysics  Science  and  Technology Centre,  carried  out  within  the  International  Research  Agendas  programme  of  the Foundation for Polish Science financed by the European Union under the European Regional Development Fund. 
This project has received funding /support from the European Union's Horizon 2020 research and innovation programme under the Marie Skłodowska-Curie grant agreement No 860881-HIDDeN. PA and MV gratefully thank AstroCeNT for their hospitality, where part of this work was developed. PA is deeply thankful to Laura Pejcha for her support and company throughout this project.

\appendix

\section{\boldmath Solutions for the $\phi-$radiation system} \label{app:phi-r}
Let us start by solving the evolution Eqs.~(\ref{eq:cosmo2}) and (\ref{eq:cosmo3}) during the dominance of $\phi$, taking into account the decays. The Hubble parameter can be written as $H\simeq \sqrt{\rho_\phi/(3M_P^2)}$. By changing variables from time to scale factor, and defining $\Heq \equiv H(\Teq)$, $u(r)=\frac{\rho_\phi}{\rho_{\rm eq}} r^\beta$, and $r=R/\Req$, the evolution for the $\phi$ field now reads
\be
\frac{du}{u^{1/2}}=-\frac{\Gamma_\phi}{\Heq }\,r^{\beta/2-1}\,dr.
\ee
Whose solution to first order in $\Gamma_\phi/\Heq$ is
\be
\rho_\phi(R)= \rhoeq \left[\frac{\Gamma_\phi}{\beta\, \Heq}-\left(\frac{\Req}{R}\right)^{\beta/2}\left(1+\frac{\Gamma_{\phi}}{\Heq\,\beta}\right)\right]^2.
\label{eq:energy_phi_2}
\ee
Replacing into Eq.~(\ref{eq:cosmo3}), using $s=(\rho+p)/T$, we find
\be
\rR(R)=\rhoeq\left[\left(\frac{\Req}{R}\right)^4+\frac{2\Gamma_\phi}{(8-\beta)\Heq}\left(\frac{\Req}{R}\right)^{\beta/2}-\frac{2\Gamma_\phi\, }{(8-\beta)\Heq}\left(\frac{\Req}{R}\right)^{4}\right].
\label{eq:energy_rad_2}
\ee
For $\beta<4$, the universe is radiation dominated before $\Req$. It is  possible to get an educated guess for $\Rc$ and $\Rend$ by considering that when the decays start to influence the SM radiation, both terms in $\rho_R$ should be of the same order, meaning
\be 
\left(\frac{\Req}\Rc\right)^4\simeq \frac{2\Gamma_\phi}{(8-\beta)\Heq}\left(\frac{\Req}{\Rc}\right)^{\beta/2},
\ee
from here, we get
\be
\Rc= \Req \left( \frac{(8-\beta)}{2 } \left( \frac{\Teq}{\Tend}\right)^2 \right)^{\frac{2}{8-\beta}}.
\ee
Similarly, the decays of $\phi$ will stop, when the two terms inside the brackets of Eq.~(\ref{eq:energy_phi_2}) become comparable, thus
\be
\left(\frac{\Req}{\Rend}\right)^\beta\simeq\frac{2\Gamma_\phi}{\beta\, \Heq}\left( \frac{\Req}{\Rend}\right)^{\beta/2},
\ee
from where we find 
\be
\Rend= \Req\left(\frac{\beta }{2} \left( \frac{\Teq}{\Tend}\right)^2\right)^{2/\beta}.
\ee
Furthermore, the corresponding temperatures can be obtained from
\bea
\rhoeq\left(\frac{\Req}{\Rc}\right)^{4}\simeq \rR(\Tc),\quad \mbox{and}\quad {{\frac{\Gamma_\phi}{\Heq}=\frac{\Tend^2}{\Teq^2},}}
\label{app:eq:cond}
\eea
which result in
\be
    \Tc \simeq \Teq \left(\frac{2\Gamma_\phi\, }{\Heq(8-\beta)}\right)^{\frac{2}{8-\beta}}=\Teq\left(\frac{4}{(8-\beta)^2} \frac{\Tend^4}{\Teq^4}\right)^\frac{1}{8-\beta}.
\ee
From this analysis it can be also extracted that deep during the $\phi$ domination, the relation between temperature and scale factor is
\be
T\simeq \Teq\left[\frac{2}{8-\beta}\frac{\Tend^2}{\Teq^2} \right]^{1/4} \left(\frac{\Req}R\right)^{\beta/8}.
\ee
We emphasise the equation above tells us that for smaller $\beta$ the temperature decreases more slowly with the universe expansion and therefore it implies that the universe is bigger when reaching the temperature $\Tend$. Let us also remark that the range of Eqs.~\eqref{eq:phi_density}-\eqref{eq:T_NSC} are only valid if the fluid dominates strongly over radiation. There can be certain cosmologies where this is not the case - and still have an impact on the axion relic density - and thus, the above referred expressions have to be handled carefully.

The case $\beta=0$ should be treated separately. In this case, we start with Eq.~(\ref{eq:cosmo2}) and defining $u=\sqrt{\rho_\phi/\rhoeq}$ and $r=\ln(R/\Req)$. Thus, the equation to solve is
\be
du=-\frac{\Gamma_\phi}{2\Heq} dr,
\ee
whose solution is given by
\be
    \rho_\phi(R) \simeq \rhoeq\left(1+\frac{\Gamma_\phi}{\Heq }\ln\left(\frac{\Req}{R}\right)\right).
\label{app:phibeta0}
\ee
By replacing into the radiation equation, we get the evolution is 
\be
    \rho_R(R) \simeq \rho_{\Req}\left(\left(\frac{\Req}{R}\right)^4 + \frac{\Gamma_\phi }{4\Heq}\right).
\label{app:radbeta0}
\ee
 We can find the moment when the radiation energy density starts to be influenced by the $\phi$ field
\be
\Rc=\Req \left(\frac{4\Heq}{\Gamma_\phi }\right)^{1/4}.
\label{Rcbeta0}
\ee
Additionally the moment when the decays of the $\phi$ field are significant is represented by
\be
\Rend=\Req \times\exp\left(-\frac{\Heq}{\Gamma_\phi}\right).
\label{Rendbeta0}
\ee

\section{\boldmath Axion oscillations during $\phi$ domination}
\subsection{Oscillations in Region 1}{\label{app:reg1}}
To analyse this epoch, first we find the oscillation temperature. Since for this period it is assumed the universe is radiation dominated, the Hubble parameter is $H=\sqrt{\rho_R/(3M_P^2)}$, so the oscillation temperature is the usual found for a QCD axion, depicted in Fig.~\ref{fig:axion_std}. There are two regimes for the oscillation temperature, depending on whether the effects of temperature on the axion mass,  are important or not, Eq.~(\ref{eq:thermal_mass}), and they are the same as the standard cosmological scenario, namely
\be
    \Tosc \simeq
    \begin{dcases} 
        \left(\ma\, M_P\right)^{1/2}, &\Tosc\lesssim \TQCD.\\
        \left(\ma M_P \TQCD^4\right)^{1/6}, & \Tosc \gtrsim \TQCD
    \end{dcases}
\ee
These two temperatures intersect at the QCD epoch, which can be expressed as a particular mass of the axion, which is the same as the standard cosmological scenario, found in Eq.~(\ref{eq:m_i}), but we rewrite here, for completeness, this time forgetting about the degrees of freedom, as we are only interested in the order of magnitude, so
\be
    3\, H(\TQCD) = m_{R_1} \simeq \frac{\TQCD^2}{M_P}.
\ee
The condition for $\Rosc$ to occur in this regime can be re-written to be $T_{\rm{osc}}\gg \Teq$ and by using the expressions for $\Tosc$ we can write it as a condition for the axion mass
\be
\ma\gg
\begin{dcases} 
\frac{\Teq^2\, }{M_P}, & \ma\lesssim m_{R_1}, \\
\frac{\Teq^6}{M_P\, \TQCD^4}, & \ma\gtrsim m_{R_1},
\end{dcases}
\label{eq:R1_boundInf}
\ee
where the first equation correspond to the limit of the axion mass in the case that the temperature has significant effects on the mass of the axion (before QCD transition) and the second is the limit after the QCD transition.

The relic abundance, if the oscillation happens in this region, is given by $\Omega_a(T_0)=\Omega_a^{\rm std}(T_0)\gamma_{R_1} $, such that $\Omega_a^{\rm std}$ is the axion relic energy density today in the standard cosmology, and $\gamma_{R_1}=S_{\rm osc}/S_{\rm end}$ the entropy dilution factor. To find the latter, we take
\bea
\gamma_{R_1}&=&\frac{S_{\rm osc}}{S_{\rm end}}=\frac{S_{\rm c}}{S_{\rm end}}= \left(\frac{\Tc\, \Rc}{\Tend\, \Rend}\right)^3,\label{app:eq:gamma_R1}\\
&=& \left(\frac{ \beta^2}{4}\left(\frac{\Teq}{\Tend }\right)^{4-\beta}\right)^{-\frac{3}{\beta}}.\nonumber
\eea
Now, replacing the dilution factor in the axion relic density we find
\bea
\Omega_{R_1}=
\begin{dcases}
\Omegastdo \left(\frac{ \beta^2}{4}\right)^{-3/\beta}\left(\frac{\Tend}{\Teq}\right)^{12/\beta-3} , & \mbox{for}\,\,\, \ma\lesssim m_{R_1}\\
\OmegastdT\left(\frac{ \beta^2}{4}\right)^{-3/\beta}\left(\frac{\Tend}{\Teq}\right)^{12/\beta-3} , & \mbox{for}\,\,\, \ma\gtrsim m_{R_1},
\end{dcases}
\eea
where we have defined the axion relic density for a standard cosmology, when temperature effects on the mass are unimportant and important, respectively, as
\bea
& &\Omegastdo \equiv 5\times 10^{-11}\left(\frac{\ma}{1\, \mbox{eV}}\right)^{-3/2}\,\theta_i^2,
\label{app:eq:Omega0STD}\\
& &\OmegastdT\equiv 2.4 \times 10^{-7} \left(\frac{\ma}{1\, \mbox{eV}}\right)^{-7/6} \,\theta_i^2.
\label{app:eq:OmegaTSTD}
\eea

\subsection{Oscillations in Region 2}{\label{app:reg2}}
We start by writing the Hubble parameter as
\be
H\simeq\sqrt{\frac{\rho_\phi}{3M_P^2}}\simeq\Heq \left(\frac{\Req}R\right)^{\beta/2}=\Heq  \left(\frac{\Tosc}{\Teq}\right)^{\beta/2},
\label{HR2}
\ee
where $\Heq\equiv H(\Req)$. We find the oscillation temperature for both cases, where the temperatures effects on the mass are and are not important, as
\bea
\Tosc^{R_2}=
\begin{dcases}
\Teq \left(\frac{M_P \ma}{\Teq^2}\right)^{2/\beta}& \ma\lesssim m_{R_2},
\\
\left(\Teq^{\frac{\beta-4}{2}}M_P\, \ma\, \TQCD^4 \right)^\frac{2}{\beta+8}& \ma\gtrsim m_{R_2}.
\end{dcases}
\label{app:eq:ToscR2}
\eea
These two temperatures intersect at the axion mass at the QCD temperature, $3\, H(\TQCD)$, in this case given by
\be
m_{R_2}= \left(\frac{\TQCD}{\Teq}\right)^{\beta/2}\,\frac{\Teq^2}{M_P}.
\label{app:eq:mR2}
\ee
By requiring $\Tc\ll \Tosc\ll \Teq$ we find the mass range of the axion to have the oscillation happening in this era. For the case of constant axion mass, the requirement is $\Tc\ll \Tosc\lesssim \TQCD$ so we find
\bea
m_{R_2}\ll \ma\ll \frac{\Teq^{6}}{M_P\,\TQCD^4}.
\label{app:eq:NSC_R2_Tbound}
\eea
While for the case the oscillation happens before the QCD phase transition, it is found
\be
\frac{\Teq^2}{M_P}\left(\frac{2\,\Tend^2}{\Teq^2 (8-\beta)}\right)^{\frac{\beta}{8-\beta}}\ll \ma\ll m_{R_2}.
\label{app:eq:NSC_R2_0bound}
\ee

The axion abundance can be found by taking the expression for the energy density in the NSC, Eq.~(\ref{eq:NSC_axion_relic_density}), where $S_{\rm osc}/S_{\rm end}$ is the entropy dilution, which is the same found in Eq.~(\ref{eq:gamma_R1}). We get
\be
\Omega_{R_2}\simeq
\begin{dcases}
 \Omegastdo \left(\frac{\Tend^2}{\ma M_P} \right)^{\frac{3}{2\beta} (4-\beta)} & \text{for } \ma\lesssim m_{R_2}\,, \\
 \OmegastdT \left(\frac{\beta^2}{4} \right)^{-3/\beta} \left(\frac{\Tend}{\Teq}\right)^{\frac{3}\beta (4-\beta)} \left[ \frac{\Teq^7}{\left(\TQCD^4 \, \ma M_P\right)^{7/6}}\right]^{\frac{4-\beta}{\beta+8}} & \text{for } \ma\gtrsim m_{R_2}\,.
 \end{dcases}
 \label{app:eq:R2_relic}
\ee
Note that in the case of a constant axion mass, the dependence on $\Teq$ cancels.

\subsection{Oscillations in Region 3}{\label{app:reg3}}

We will assume $\Rosc$ is still far from $\Rend$, such that the Hubble parameter, for our purposes,  can still be considered as
\be
H\sim \sqrt{\frac{\rho_\phi}{3\, M_P^2}}=\Heq \left(\frac{\Req}R\right)^{\beta/2}.
\ee
Nonetheless, the decays of $\phi$ are affecting radiation, such that its energy density it is better described as
\be
\rho_R \sim \rhoeq\,\frac{2\Gamma_\phi}{(8-\beta) \Heq}\left(\frac{\Req}R\right)^{\beta/2}.
\ee
Now we can equate the above expression with the energy density as a function of the temperature, the well known $\rho_R=(\pi^2/30) \gs(T) T^4$, such that we can obtain the relation between scale factor and temperature in this region to be
\be
\left(\frac{\Req}R\right)^{\beta/2}=\left( \frac{8-\beta}{2}\right) \frac{T^4}{ \Tend^2 \Teq^2}.\label{eq:R_vs_T_3}
\ee
Replacing back into the Hubble parameter, we find\footnote{Dropping the degrees of freedom.}

\be
H(T)=\Heq \left(\frac{8-\beta}2\right)\frac{T^4}{\Tend^2\Teq^2}.
\ee
As in the previous regions, we can find the oscillation temperature
\bea
\Toscthree=
\begin{dcases}
\left(\frac{2 \ma M_P\, \Tend^2}{8-\beta} \right)^{1/4} & \mbox{for}\,\,\, \ma\lesssim m_{R_3}\\
\left(\frac{2 \ma M_P\, \TQCD^4\, \Tend^2}{8-\beta}\right)^{1/8}& \mbox{for}\,\,\, \ma\gtrsim m_{R_3}.
\end{dcases}
\label{app:eq:ToscR3}
\eea
So for this region the dependence of $\Tosc$ on $\beta$ is quite mild, and on $\Teq$ nonexistent.  
The intersection between these temperatures gives the moment of the QCD phase transition, which can be used to obtain the mass of the axion at this point. We get
\be
m_{R_3}=\frac{(8-\beta)}2\frac{\TQCD^4}{M_P \,\Tend^2}\simeq 2\times 10^{-13}\, \mbox{eV} \left(\frac{\mbox{GeV}}{\Tend}\right)^2.
\label{app:eq:mass_R3}
\ee
We now can write down the mass ranges expected for the oscillation of the axion field to happen in region 3. Here it is better to work with the requirement $\Rc\ll R_{\rm osc}\ll \Rend$, but it can be put into temperatures  by using Eq.~(\ref{eq:R_vs_T_3}). Firstly, let us assume the oscillation happens completely before the QCD transition, where temperature effects on the axion mass are important. In that case 
\be
m_{R_3}\ll \ma\ll \frac{\Teq^6 }{M_P\,\TQCD^4}\left(\frac{2\,\Tend^2 }{(8-\beta)\Teq^2}\right)^{\frac{8+\beta}{8-\beta}},
\label{eq:mass_range_R3_T}
\ee
On the other hand, for the constant axion mass regime, the requirement maps into a mass range of as
\be
\frac{2}{\beta} \frac{\Tend^2}{M_P}\ll \ma \ll \frac{\Teq^2 }{M_P }\left(\frac{2\,\Tend^2 }{(8-\beta)\Teq^2}\right)^{\frac{\beta}{8-\beta}}.
\ee
The dilution factor, $S_{\rm osc}/S_{\rm end}$, is different in the previous regions, because the axion is produced in an era where the SM entropy is increasing. We write
\be
\gamma_{R_3}=\frac{S_{\rm osc}}{S_{\rm end}}=\left(\frac{\Tosc R_{\text{osc}}}{\Tend\Rend}\right)^3.
\ee
From Eqs.~\eqref{eq:R_vs_T_3} and~\eqref{eq:rend} we can easily find the ratio $\Rosc/\Rend$ to be
\be
\frac{\Rosc}{\Rend}=\left[\frac{4}{\beta(8-\beta)}\left(\frac{\Tend}{\Tosc}\right)^4\right]^{2/\beta},
\ee
and then the dilution factor is 
\be
\gamma_{R_3}=\left(\frac{4}{\beta(8-\beta)}\right)^{6/\beta} \left(\frac{\Tend}{\Tosc}\right)^{24/\beta-3}.
\ee
Now the axion abundance is easily found 
\bea
\Omega_{R_3}=\begin{dcases}
 \Omegastdo\left(\frac{2}{\beta}\right)^{6/\beta} \left(\frac{\Tend^2}{\ma M_P}\right)^{\frac{3}{2\beta} (4-\beta)} & \mbox{for}\,\, \ma\lesssim m_{R_3}\,,\\
\OmegastdT\, (8-\beta)^{\frac{\beta+6}{2\beta}} \left(\frac{\Tend^6}{\TQCD^4 \ma M_P}\right)^{3/\beta-2/3} & \mbox{for}\,\, \ma\gtrsim m_{R_3}\,.
\end{dcases}
\label{app:eq:R3_relic}
\eea
Both expressions are independent on $\Teq$, showing that for this region the main parameters are $\Tend$ and $\beta$.

\section{Adiabatic invariant and the anharmonic factor}\label{app:anF}
In oscillatory systems with varying period, the energy is not conserved, and it is usually useful to define an ``adiabatic invariant",
\begin{equation}
	J \equiv \oint p \ d \theta \;,
	\label{eq:adiabatic_inv_def}
\end{equation}
which is an approximate constant of motion.

To calculate the adiabatic invariant of the axion, we note that the Hamiltonian that results in the EOM of Eq.~(\ref{axion_eom}) is
\begin{equation}
	\mathcal{H} = \dfrac{1}{2} \dfrac{p^2}{\fa^2 \ R^3} + V(\theta) \ R^3\;,
	\label{eq:axion_H}
\end{equation}
with 
\begin{eqnarray}
	& p = \fa^2 \ R^3 \ \dot \theta \\
	\label{eq:momentum}
	& V(\theta) = \maT^2(t)\, \fa^2\, (1-\cos \theta) \;.
	\label{eq:potential}
\end{eqnarray}

\begin{figure}[t]
	\includegraphics[width=1\textwidth]{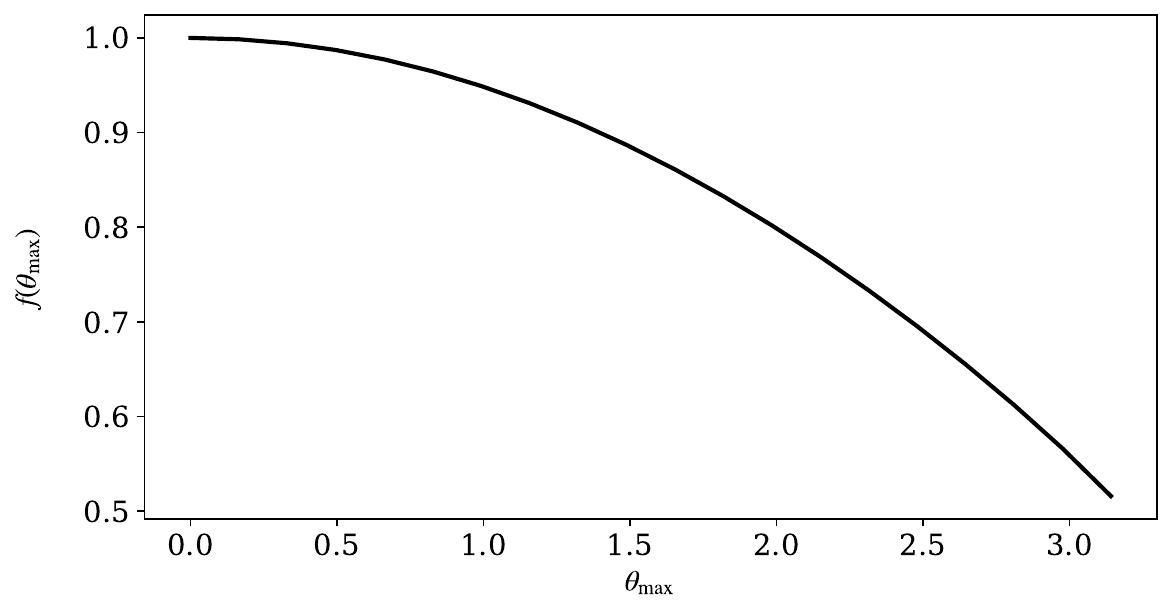}
	\caption{The anharmonic factor for $0 \leq \theta_{\rm max}< \pi $.}
	\label{fig:anharmonic_factor}
\end{figure}

Notice that the explicit time dependence comes from $R$ and $\maT$. That is, the Hamiltonian varies slowly if $\dot{\tilde{m}}_a/\maT \ll \maT$ and $H \ll \maT$, which are the adiabatic conditions.   When these conditions are met, the adiabatic invariant of this system becomes
\begin{equation}
	J =2\sqrt{2} \ \fa^2 \ \maT(t) \, R^3 \ \int_{- \theta_{\rm max}}^{\theta_{\rm max}} \sqrt{\cos \theta - \cos \theta_{\rm max}} \ d \theta  
	\;,
	\label{eq:J_axion_derivation}
\end{equation}
where we have defined $\theta_{\rm max}$ the maximum $\theta$ during its oscillation, which corresponds to $p=0$. Moreover, we have assumed negligible change of $\maT$ and $R$ during one period, due to the adiabatic conditions.
For the sake of consistency with the literature~\cite{Lyth:1991ub, Bae:2008ue}, we define $I \equiv J/(\pi \ \fa^2)$; a rescaled adiabatic invariant 
\begin{equation}
	I \equiv R^3 \ \maT(t) \ \theta_{\rm max}^2  \, f(\theta_{\rm max})  \;,
	\label{eq:J_axion_def}
\end{equation}
where 
\begin{equation}
	f(\theta_{\rm max}) =\dfrac{ 2 \sqrt{2}}{\pi \theta_{\rm max}^2 } \int_{- \theta_{\rm max}}^{\theta_{\rm max}} d \theta \sqrt{ \cos \theta - \cos \theta_{\rm max} } \;,
	\label{eq:anharmonic_f}
\end{equation}
is called the anharmonic factor, with $ 0.5 \lesssim f(\theta_{\rm max}) \leq 1$, as shown in Fig.~\ref{fig:anharmonic_factor}. 
It should be noted that, although the anharmonic factor tends to decrease for large angles, the overall relic abundance is enhanced as the initial angle increases. This effect is due to an ``initial anharmonic correction", discussed in detail in Ref.~\cite{Bae:2008ue}, which causes the relic abundance to increase significantly as $\theta \to \pi$.
In this article, this effect can be seen by comparing the two curves in the right panel of Fig.~\ref{fig:axion_std}, where our analytical estimate for the initial misalignment angle is larger than the numerical value.

The adiabatic invariant allows us to calculate the maximum value of the angle $\theta$ at late times from its corresponding value at some point after the adiabatic conditions where met (at some $T \ll \Tosc$). Numerically, this means that one can stop the integration once the axion starts to evolve adiabatically -- at some temperature $T_{\rm ad}$ with $\theta=\theta_{\rm ad}$. Then, since $J$ is conserved between $T_{\rm ad}$ and $T_0$,  the axion angle at $T=T_0$ becomes
\begin{equation}
    \theta_0^2 = \dfrac{s_0}{s_{\rm ad}} 
    \dfrac{\tilde m_{a, {\rm ad}}}{\ma}\theta_{\rm ad}^2 \ f(\theta_{\rm ad}) \times  \dfrac{S_{\rm ad}}{S_0} \;,
    \label{eq:theta_0_anF}
\end{equation}
where the subscript ``${\rm ad}$" indicates that the corresponding quantity is evaluated at $T=T_{\rm ad}$.
Therefore, the energy density of the axion becomes
\begin{equation}
    \rho_a(T_0) = \dfrac{1}{2} \ \dfrac{s_0}{s_{\rm ad}} 
    \ \fa \ \ma \ \tilde m_{a, {\rm ad}} \ \theta_{\rm ad}^2 \ f(\theta_{\rm ad}) \times  \dfrac{S_{\rm ad}}{S_0} \;.
    \label{eq:rho_a0_anF}
\end{equation}
Although this form of the axion energy density is similar to  Eq.~(\ref{eq:NSC_axion_relic_density}) -- with $\Tosc \to T_{\rm ad}$,  multiplied with the anharmonic factor in Eq.~\eqref{eq:anharmonic_f}.  It is valid for any value of $\theta_i$, as it takes into account the exact form of the potential of the axion field.

\bibliographystyle{JHEP}
\bibliography{biblio}

\end{document}